\documentclass{IEEEtran}
\usepackage[utf8]{inputenc}

\usepackage{psfrag,epsfig,graphics}
\usepackage{amsmath,amsthm,amssymb,multirow}
\usepackage{mathbbol}
\usepackage{amssymb}             %

\DeclareSymbolFontAlphabet{\amsmathbb}{AMSb}%
\usepackage{mathabx} %
\usepackage{mathrsfs}
\usepackage{cuted}
\usepackage{adjustbox}

\usepackage{enumitem}

\usepackage[font={footnotesize}]{caption, subfig}

\usepackage{url}

\usepackage{booktabs}
\usepackage{graphicx}

\usepackage[noadjust]{cite}
\usepackage{multirow}
\usepackage[lined,linesnumbered,ruled]{algorithm2e}

\usepackage{color}  %

\def\cblue{\textcolor{blue}}

\newcommand{\cb}[1]{{\boldsymbol{#1}}}
\newcommand{\cp}[1]{\ifmmode {\mathcal{#1}}\else ${\mathcal{#1}}$\fi}

\newcommand{\bA}{\boldsymbol{A}}

\newcommand{\bD}{\boldsymbol{D}}

\newcommand{\bI}{\boldsymbol{I}}

\newcommand{\bM}{\boldsymbol{M}}

\newcommand{\bV}{\boldsymbol{V}}
\newcommand{\bW}{\boldsymbol{W}}

\newcommand{\bY}{\boldsymbol{Y}}

\newcommand{\ba}{\boldsymbol{a}}

\newcommand{\bc}{\boldsymbol{c}}

\newcommand{\bm}{\boldsymbol{m}}

\newcommand{\bh}{\boldsymbol{h}}

\newcommand{\br}{\boldsymbol{r}}
\newcommand{\by}{\boldsymbol{y}}

\newcommand{\bx}{\boldsymbol{x}}

\newcommand{\bnu}{\boldsymbol{\nu}}

\newcommand{\calL}{\mathcal{L}}
\newcommand{\calN}{\mathcal{N}}

\newcommand{\calQ}{\mathcal{Q}}

\usepackage[euler]{textgreek}
\usepackage[colorinlistoftodos]{todonotes}

\newcommand{\balpha}{\boldsymbol{\alpha}}

\newcommand{\bgamma}{\boldsymbol{\gamma}}
\newcommand{\bepsilon}{\boldsymbol{\epsilon}}
\newcommand{\bpsi}{\boldsymbol{\psi}}

\newcommand{\bpi}{\boldsymbol{\pi}}
\newcommand{\bmu}{\boldsymbol{\mu}}
\newcommand{\bsigma}{\boldsymbol{\sigma}}

\newcommand{\bxi}{\boldsymbol{\xi}}
\newcommand{\bzeta}{\boldsymbol{\zeta}}

\newcommand{\bPsi}{\boldsymbol{\Psi}}
\newcommand{\bSigma}{\boldsymbol{\Sigma}}

\newcommand{\bUpsilon}{\boldsymbol{\Upsilon}}

\newcommand{\tr}{\operatorname{tr}}
\newcommand{\Ex}{\amsmathbb{E}}
\newcommand{\vect}{\operatorname{vec}}
\newcommand{\diag}{\operatorname{diag}}

\newcommand{\Dir}{\operatorname{Dir}}
\newcommand{\KL}{\operatorname{KL}}

\newcommand{\forw}{\text{forw}}
\newcommand{\back}{\text{back}}

\usepackage{color}  %

\def\cblue{}

\definecolor{darkgreen}{rgb}{0., 0.4, 0.}
\definecolor{amber}{rgb}{1.0, 0.49, 0.0}
\definecolor{orange}{rgb}{1.0, 0.4, 0.0}

\usepackage{tikz}
\usetikzlibrary{bayesnet}
\usetikzlibrary{arrows}
\usetikzlibrary{positioning}

\title{Dynamical Hyperspectral Unmixing with Variational Recurrent Neural Networks}

\author{Ricardo A. Borsoi, \IEEEmembership{Member,~IEEE}, Tales Imbiriba, Pau Closas, \IEEEmembership{Senior Member,~IEEE} 
\thanks{R.A. Borsoi is with the University of Lorraine, CNRS, CRAN, Nancy, F-54000, France (e-mail: \mbox{\{ricardo.borsoi\}@univ-lorraine.fr}).}
\thanks{T. Imbiriba and P. Closas are with the Dept. of Electrical \& Computer Engineering, Northeastern University, Boston, MA 02115, USA (e-mail: \mbox{\{talesim,closas\}@ece.neu.edu}).}
\thanks{This work has been partially supported by the NSF under Award ECCS-1845833.}
}

\allowdisplaybreaks
\begin{document}

\maketitle
\begin{abstract}
Multitemporal hyperspectral unmixing (MTHU) is a fundamental tool in the analysis of hyperspectral image sequences. It reveals the dynamical evolution of the materials (endmembers) and of their proportions (abundances) in a given scene.
However, adequately accounting for the spatial and temporal variability of the endmembers in MTHU is challenging, and has not been fully addressed so far in unsupervised frameworks.
In this work, we propose an unsupervised MTHU algorithm based on variational recurrent neural networks.
First, a stochastic model is proposed to represent both the dynamical evolution of the endmembers and their abundances, as well as the mixing process. Moreover, a new model based on a low-dimensional parametrization is used to represent spatial and temporal endmember variability, significantly reducing the amount of variables to be estimated.
We propose to formulate MTHU as a Bayesian inference problem. However, the solution to this problem does not have an analytical solution due to the nonlinearity and non-Gaussianity of the model. Thus, we propose a solution based on deep variational inference, in which the posterior distribution of the estimated abundances and endmembers is represented by using a combination of recurrent neural networks and a physically motivated model. The parameters of the model are learned using stochastic backpropagation.
Experimental results show that the proposed method outperforms state of the art MTHU algorithms.
\end{abstract}


\section{Introduction}
\label{sec:intro}

Hyperspectral images (HIs) have very high spectral resolution, which allows for a precise discrimination of different materials in a scene~\cite{Bioucas-Dias-2013-ID307}. However, physical limitations of spectral image acquisition and large distances between the sensor and the scene of interest as seen in, e.g., remote sensing, means that each pixel of an HI may cover a large area of the scene and typically contains a mixture of different materials~\cite{shaw2003spectralImagRemote}. 
Hyperspectral unmixing (HU) aims to decompose an HI into the spectral signatures of the pure materials it contains (the \emph{endmembers} -- EMs), and the proportions with which they appear in each pixel (the \emph{abundances})~\cite{Keshava:2002p5667}. 

The classical approach to describe the interactions between light and the different materials in a pixel is the linear mixing model (LMM)~\cite{Keshava:2002p5667}. However, the LMM assumes the EM signatures to be the same for all pixels in an HI, disregarding spectral variability of the EMs which can be caused by, e.g., atmospheric, illumination or seasonal variations, and propagates significant errors throughout the HU processing chain~\cite{borsoi2020variabilityReview2,Zare-2014-ID324-variabilityReview}. Thus, significant effort has been dedicated to address spectral variability in HU (see Section~\ref{sec:relatedw} for a review).

More recently, multitemporal HU (MTHU) has been receiving increasing interest in the literature since it leverages information in sequences of HIs acquired at different time instants to reveal the dynamical evolution of the endmembers and abundances in a scene~\cite{henrot2016dynamical,thouvenin2016online,borsoi2020multitemporalUKalmanEM,liu2021bayesianSU_multitemporal}. 
MTHU has proven important for many applications 
such as invasive species mapping in rainforests~\cite{somers2013invasiveHawaiiMultiTemporalBandWeighting,somers2013uncorrelatedBandSelectionInstabilityIndex}, and monitoring vegetation cover in shrublands~\cite{lippitt2018multidateMESMAshrublands} or seasonal variations of vegetation cover in dry forests~\cite{goenaga2013unmixingTimeSeriesPuertoRico}. Moreover, MTHU is also useful to perform change detection at the subpixel level~\cite{liu2019reviewCD,guo2021changeDetUnmixing}.
However, spectral variability can be very significant in MTHU due to different seasonal and acquisition conditions~\cite{thouvenin2016online,borsoi2020variabilityReview2,Zare-2014-ID324-variabilityReview}.

Addressing both the spatial and temporal spectral variability of the EMs is challenging, and has only been done in MTHU by supervised techniques~\cite{somers2013invasiveHawaiiMultiTemporalBandWeighting,borsoi2021MT_MESMA}. However, supervised MTHU techniques require prior knowledge of libraries containing spectral signatures which can accurately represent the endmembers for each image in the time sequence. Such libraries can be difficult or expensive to collect. Unsupervised MTHU methods, on the other hand, estimate both the endmember signatures and the abundances for all time instants directly from the observed HI sequence. Thus, unsupervised methods are of great practical interest, but can be challenging to design. See Section~\ref{sec:relatedw} for a review of MTHU methods.

Machine learning has become a popular framework to solve the HU problem~\cite{bhatt2020deepLearningHUreview}.
Recent developments include methods based on, e.g., autoencoders (AECs)~\cite{palsson2022unmixingAECcomparison} or unrolled optimization-based neural networks~\cite{zhou2021ADMM_SU_networks} (see Section~\ref{sec:relatedw} for a review). 
In particular, solutions based on deep learning are especially attractive for HU when the mixing model considers nonidealities such as, e.g., nonlinearity~\cite{wang2019AECnlin} and EM variability~\cite{borsoi2019deepGun}, circumventing the need to construct complex analytical models to represent such physical effects.

However, the literature lacks MTHU solutions that are unsupervised and take spatial and temporal EM variability into account, which are addressed in this work.
In particular, we also address several other needs, including the development of parsimonious models for EM variability with adjustable flexibility, and of machine learning-based strategies for MTHU which jointly leverage both a physically motivated and data-driven (e.g., neural networks) models in a principled manner. Such hybrid approaches, where physics-informed models are used to regularize and provide interpretability to data-driven methodologies are becoming increasingly popular~\cite{imbiriba2022hybrid}.

In this work, we propose an unsupervised MTHU algorithm based on variational recurrent neural networks (RNNs). 
First, a stochastic model is proposed to represent both the dynamical evolution of the EMs and of the abundances, as well as the mixing process. 
Moreover, a new low-dimensional model is used to represent spatial and temporal EM variability by parametrizing band-dependent scaling variations of the EMs using a set of smooth spectral basis functions. This allows us to control the flexibility of the model by varying the number and types of basis functions.
To model the abundances, we make use of the \emph{softmax basis} representation~\cite{barber2012ML_book}, which leads to a physically accurate model and has been successfully used for fuzzy classification~\cite{kent1988spatialClassificationFuzzy} and single-image HU~\cite{eches2011bayesianSpatialMarkovUnmixing}. This way, we can use a Gaussian distribution to represent the abundances in the softmax basis, which closely approximates a Dirichlet distribution when mapped back to the original abundance domain (i.e., the unit simplex)~\cite{mackay1998choiceBasisDirichlet}.

MTHU is then formulated as a Bayesian inference problem. However, exact inference is analytically intractable due to the nonlinearities in the model. Thus, we consider a variational inference solution based on RNNs, in which the approximate joint posterior distribution of abundances and EMs is learned by maximizing a lower bound over the marginal likelihood of all pixels. 
Note that approximating the true posterior typically requires a flexible family of distributions which can be represented using neural networks~\cite{kingma14VAEs}. However, using feedforward neural networks leads to models with large numbers of parameters, making inference costly. By exploiting the temporal structure in the data (e.g., Markovity), RNNs provide a solution that gives flexibility while also having a lower number of parameters (being computationally lightweight). Besides, RNNs have shown excellent performance in numerous sequence modeling tasks~\cite{lipton2015criticalReviewRNNs}.
Interpretability of the estimated abundances and EMs is paramount for the applicability of MTHU systems. For this reason, we parameterized the joint posterior distribution using a hybrid model composed of physics-based and data-driven components. More specifically, the posterior is modeled by a family of nonlinear functions constructed by integrating both a simple, physically motivated model that is able to provide an approximate abundance estimate, and a bidirectional RNN that can represent more complex effects (i.e. not captured by the simpler model).
The parameters of the model and of the posterior distribution are learned based on all image pixels using stochastic gradient descent (SGD).
The contributions of this paper include:
\begin{itemize}[partopsep=0pt, topsep=-\parskip, parsep=0pt, itemsep=0pt] %
    \item a new low-dimensional model to represent the spatial and temporal variability of the EMs with a small amount of parameters to be learned;
    \item a stochastic model describing the temporal evolution of the abundances and of the EM variability parameters, which leverages the softmax basis used in single-image HU~\cite{eches2011bayesianSpatialMarkovUnmixing} to obtain a physically accurate representation of the abundance dynamics using a Gaussian distribution;
    \item a deep variational inference formulation of model-based MTHU with both spatial and temporal EM variability, solved using stochastic backpropagation;
    \item a parametrization of the posterior distribution of the abundances and EMs combining bidirectional RNNs and a physically interpretable model.
\end{itemize}
The proposed method is called ReDSUNN for \textit{Recurrent hyperSpectral Unmixing with Neural Networks}.
Experimental results with synthetic and real data show that ReDSUNN outperforms state of the art MTHU algorithms. Codes are available at \url{https://github.com/ricardoborsoi/ReDSUNN}.

\begin{figure*}
    \centering
    \includegraphics[width=1\linewidth]{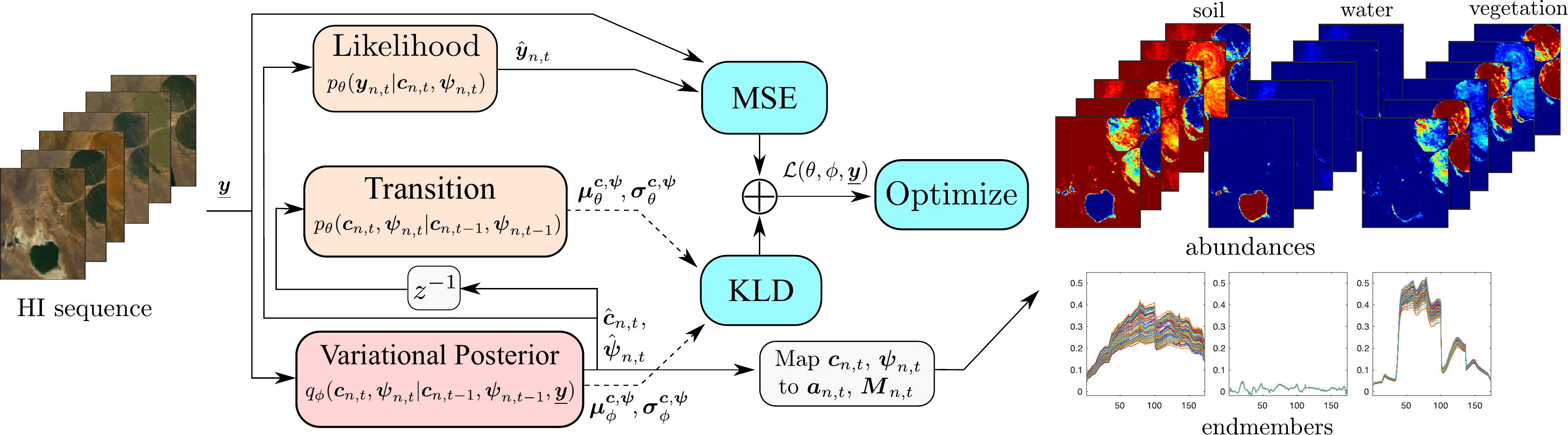}
    \vspace{-3ex}
    \caption{Illustrative diagram of the proposed ReDSUNN method. A likelihood and transition PDFs, with parameters $\theta$, define the spatiotemporal mixing process (i.e., the generative model). The variational posterior, with parameters $\phi$, approximates the unmixing solution (i.e., the PDF of the abundances and EMs conditioned on the HIs) recursively, being implemented using an RNN. The parameters of these PDFs are learned \cblue{by maximizing} a loss function based on the ELBO, which balances data reconstruction (MSE) and consistency between posterior and prior (KLD). The parameters $\bc_{n,t}$ and $\bpsi_{n,t}$ are mapped to the abundances and EM matrices, $\ba_{n,t}$ and $\bM_{n,t}$ using a deterministic model. The notation $z^{-1}$ represents a delay. See Section~\ref{sec:methodOverview} for more details.}
    \label{fig:overall_diagram}
    \vspace{-1.5ex}
\end{figure*}

\begin{figure}[htb]
    \centering
         \includegraphics[width=0.38\linewidth]{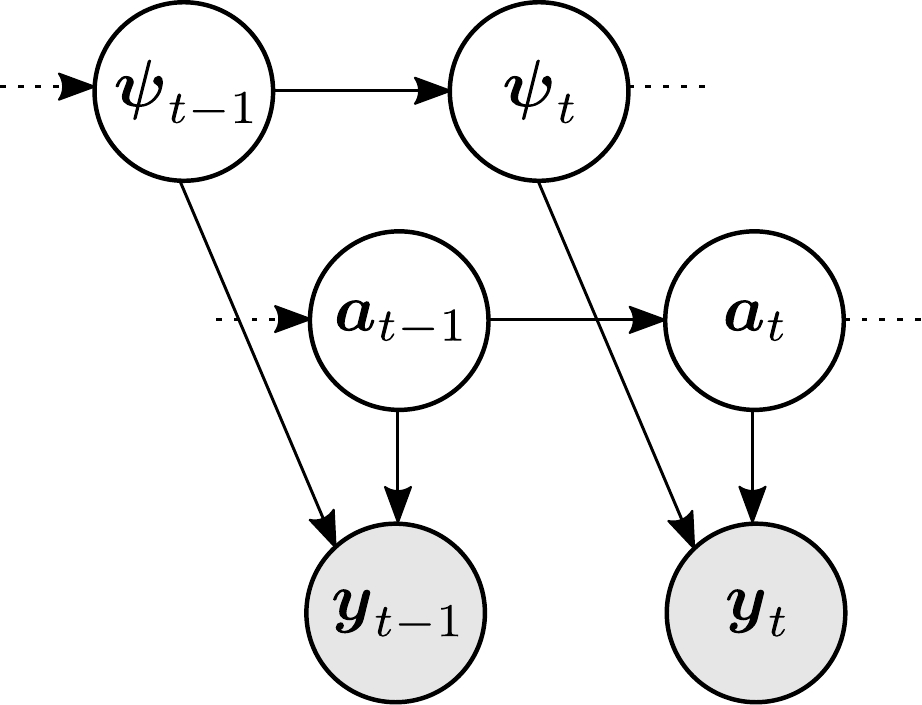}
    \qquad
         \includegraphics[width=0.3\linewidth]{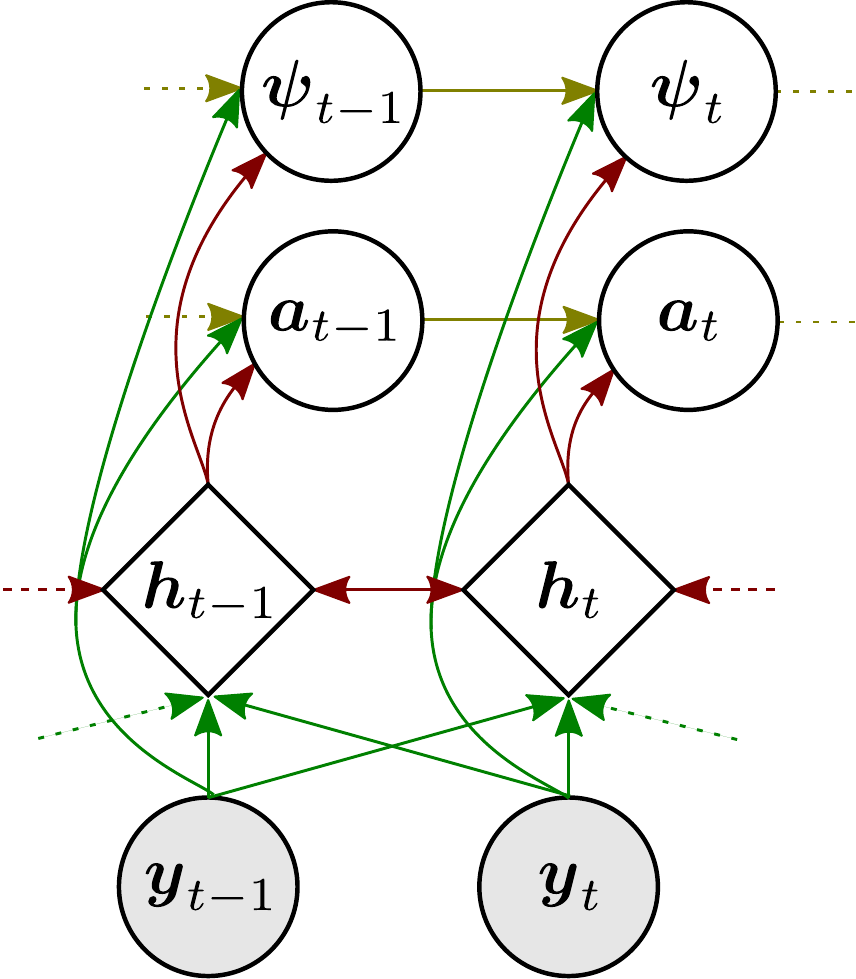}
    \vspace{-1.5ex}
    \caption{%
    In the generative model (left), conditional independence among pixels  $\by_t$ is assumed given abundances $\ba_t$ and variability coefficients $\bpsi_t$, as well as Markovity of state variables. During inference (right), state estimation exploits correlations between pixels, states and hidden representations $\bh_t$ generated by a bi-directional RNN.}
    \label{fig:graphical_models}
    \vspace{-0.3cm}
\end{figure}

\section{Background and related work}
\label{sec:relatedw}

A general multitemporal linear mixing model represents the $n$-th pixel of an HI acquired at time $t$ as:
\begin{align} \label{eq:LMM}
\begin{split}
    & \by_{n,t} = \bM_{n,t} \ba_{n,t} + \br_{n,t} \,, 
    \\&
    \text{s.t.}\,\,\,\,\cb{1}^\top\ba_{n,t} = 1,\,\,\, \ba_{n,t} \geq \cb{0} \,,%
\end{split}
\end{align}
where, for each time $t\in\{1,\ldots,T\}$ and pixel $n\in\{1,\ldots,N\}$, $\by_{n,t}\in\amsmathbb{R}^{L}$ denotes the observed pixel with $L$ bands, the columns of $\bM_{n,t}\in\amsmathbb{R}^{L\times P}$ contain the spectral signatures of the $P$ endmembers in the scene, vector $\ba_{n,t}\in\amsmathbb{R}^P$ contains the fractional abundances of each EM, and $\br_{n,t}\in\amsmathbb{R}^{L}$ represents additive noise. Note that the general model~\eqref{eq:LMM} can accommodate EM variability both in space and in time, being able to represent effects such as, e.g., atmospheric, illumination, or seasonal variations~\cite{borsoi2020variabilityReview2}.
In the following, we review different HU strategies addressing multitemporal sequences, spatial EM variability, and based on deep learning frameworks.

\subsection{Multitemporal HU}

A fundamental aspect of MTHU is taking into account the relationship between the EMs and abundances at different time instants. Since these are usually temporally correlated, this can greatly improve the performance of unmixing algorithms.
Most previous works have been focused on addressing the variability of the EMs in time. This was usually performed by considering a more constrained version of model~\eqref{eq:LMM}, where only the temporal variability of the EMs is considered, leading to $\bM_{n,t}=\bM_{m,t}$, for all $1\leq n,m\leq N$ \cite{henrot2016dynamical,thouvenin2016online,borsoi2020multitemporalUKalmanEM}.

Several works have considered parametric models to represent \cblue{the} temporal variability of the EMs (i.e., using only a single EM matrix per image). For instance, dynamical model was used in~\cite{henrot2016dynamical} to constrained the EMs to be a scaled versions of a reference EM matrix, with smoothly varying scaling factors. Another model considered the EM matrices at each time instant to be an additive perturbation of a mean EM matrix~\cite{thouvenin2016online}. %
Using this model, MTHU was performed using a two-stage stochastic programming approach~\cite{thouvenin2016online}. Other works considering this model have also proposed MTHU solutions using a distributed algorithm using sparsity constraints~\cite{sigurdsson2017sparseDistU}, and a hierarchical Bayesian framework incorporating additive residual terms to account for outliers~\cite{thouvenin2018hierarchicalBU}.
A recent work proposed a hierarchical Bayesian MTHU strategy (called HBUN) which incorporated priors promoting the spectral smoothness of the estimated EM signatures, and the spatial and temporal smoothness of the abundances~\cite{liu2021bayesianSU_multitemporal}. 
Another model representing the EMs using bandwise multiplicative scalings of a set of reference EM signatures was considered in~\cite{borsoi2020multitemporalUKalmanEM}. MTHU was performed by combining a Bayesian filtering (the Kalman filter and smoother) with the expectation maximization method.
Heteroscedastic measurement noise was also considered in~\cite{zhuo2022spectralTemporalBayesianSUSentinel2}, where the (diagonal) covariance matrix of the measurement noise was estimated along with the EMs and abundances in a \textit{maximum a posteriori} framework.

However, these methods disregard EM variability within a single image, which limits their performance. Existing MTHU algorithms which tackle both spatial and temporal variabilty are based on the MESMA framework~\cite{roberts1998originalMESMA}, \cblue{which} searches for a combination of EM signatures in a library which can best reconstruct a given HI, rendering it highly interpretable. These have been applied for MTHU in vegetation monitoring applications~\cite{somers2013invasiveHawaiiMultiTemporalBandWeighting,lippitt2018multidateMESMAshrublands,dudley2015multitemporalLibraryPhenologicalGradiantsMESMA}. Recent advances were also made on developing efficient solutions to this problem with theoretical guarantees by exploring the temporal information of the abundances~\cite{borsoi2021MT_MESMA}.
However, MESMA-based approaches are computationally costly and their performance depends strongly on the accuracy of the library, rendering this model inadequate for unsupervised processing.
Note that some works proposed to leverage complementary high-resolution (Landsat) \cblue{images} to unmix low-resolution (MODIS) images~\cite{wang2021spatioTemporalSUtimeseries,wang2021realTimeSU_MODIS}. However, the availability of such complementary information and difficulties associated to differences in their acquisition times limit the applicability of such methods in practice.

\subsection{HU with EM variability}

The variability of the EMs across an HI occurs due to atmospheric, illumination (e.g., topographic) or intrinsic variations of the EM spectra \cite{borsoi2020variabilityReview2,somers2011variabilityReview,Zare-2014-ID324-variabilityReview}. It introduces errors which propagate through the various steps of the traditional HU processing chains and can have significant negative impact on the abundance estimation performance. EM variability is generally addressed in HU by representing the spectral signature of each material using structured spectral libraries, statistical distributions, or physically motivated parametric models~\cite{borsoi2020variabilityReview2}.

Spectral library-based methods represent the EMs in each pixel as one of several spectral signatures in a dictionary known a priori, which implies formulating HU as a structured sparse regression problem. The sparsity prior can be addressed either with combinatorial approaches, which are computationally costly~\cite{roberts1998originalMESMA}, or using different relaxations of the L$_0$ seminorm based on convex~\cite{iordache2011sunsal,borsoi2018superpixels1_sparseU} or non-convex~\cite{drumetz2019SU_bundlesGroupSparsityMixedNorms} sparsity promoting penalties. Other strategies also allow each EM to be a convex combination of library spectra~\cite{uezato2018SU_variabilityAdaptiveBundlesDoubleSparse}. Such relaxations are computationally easier to solve.
This can be a reasonable modeling assumption when multiple signatures of the same EM can contribute to form a single pixel. However, such \cblue{relaxations might reduce the interpretability} of the solutions if the goal is to identify which signature is active~\cite{drumetz2019SU_bundlesGroupSparsityMixedNorms}.
Nevertheless, the performance of both kinds of strategies is strongly dependent on the quality of the spectral library.

Using statistical distributions to model the endmembers has been well-investigated as it provides principled HU solutions through a Bayesian framework. The Gaussian~\cite{halimi2015unsupervisedBayesianUnmixing}, mixture of Gaussians~\cite{zhou2018variabilityGaussianMixtureModel} or Beta~\cite{du2014spatialBetaCompositional} distributions have been considered.
However, when complex distributions (such as the Beta or mixtures of Gaussians) are used to represent the endmembers, HU (which consists in a Bayesian inference problem) can become computationally expensive.

Another approach to address spectral variability consists in representing the signatures of the EMs in each pixel using a physically meaningful parametric model, and estimating the model parameters during HU. Examples of such models include the use of additive perturbations~\cite{Thouvenin_IEEE_TSP_2016_PLMM}, spectrally uniform or spectrally localized multiplicative scaling factors~\cite{drumetz2016blindUnmixingELMM,imbiriba2018GLMM}, or a combination thereof~\cite{hong2019augmentedLMMvariability}. Other models explicitly exploit spatial information, using multiscale~\cite{Borsoi_multiscaleVar_2018_TIP} or low-rank tensor representation~\cite{imbiriba2018ULTRA_V}, and external information (e.g., LiDAR) by means of digital surface models~\cite{uezato2020illuminationSU_surfaceModel}. However, designing physically accurate models whose parameters can at the same time be properly recovered from an HI can be challenging.

\subsection{Deep learning-based HU}

Deep learning has recently become a popular approach to perform HU. While HU was traditionally viewed as a regression problem (i.e., learning a mapping from the pixels to the abundances)~\cite{guilfoyle2001comparativeUnmixingNeuralNetworksRBF,plaza2010selectingTrainingSamplesNNunmixing}, recent work has been focused on developing unsupervised or self-supervised strategies, which avoid the need for vast amounts of training data.
Among such strategies, AECs have become a predominant approach for deep learning-based HU due to their close connection to linear or nonlinear mixing models and good experimental performance~\cite{palsson2022unmixingAECcomparison}. The latent representations of the image pixels obtained by the AEC are associated to the abundances, and the decoder network to the mixing model~\cite{guo2015autoencodersUnmixing,palsson2018autoencoderUnmixing_IEEEaccess}.
Several AEC methods for HU have been proposed for linear HU by using different choices of encoder networks including, e.g., denoising layers~\cite{su2019deepAutoencoderUnmixing,sahoo2022HU_geological_latentEncoding}, spatial-spectral (convolutional) architectures~\cite{palsson2020convolutionalAEC_SU,hua2021gatedAEC_SU} and the use of sparsity constraints~\cite{qu2018udas_autoencoderUnmixing}. 

AECs have also been used to perform nonlinear HU by considering nonlinear decoder networks to address complex mixing effects. This includes a post-nonlinear mixing model~\cite{wang2019AECnlin}, additive nonlinearities~\cite{zhao2021AECnonlinearSUattitive3D} and the use of application-specific nonlinear neural network layers~\cite{shahid2021unsupervisedSUautoencoder}. Another work also exploited the relationship between the encoder and decoder networks to propose a model-based architecture~\cite{li2021modelBasedAECsSU}.

Spectral variability was also addressed using deep learning methods. In~\cite{borsoi2019deepGun}, a generative EM model is proposed to represent the variability of the EMs on a low-dimensional manifold. Such a model was used to perform HU using methods inspired by matrix factorization~\cite{borsoi2019deepGun}, sparse regression~\cite{borsoi2019EMlibManInterpVAE} and probabilistic approaches~\cite{shi2021generativeModelEMvariability,sahoo2022HU_geological_latentEncoding}. Gaussian process regression has also been considered as a non-parametric approach to mitigate the effects of spectral variability in HU~\cite{uezato2016unmixingGaussianProcessVariability,koirala2020geodesicSupervisedSUvariability}.

Other approaches also used multiple sets of pure pixels extracted in a self-supervised manner to regularize (explicitly or implicitly) AEC-based HU algorithms in order to improve their robustness~\cite{jin2021two_stream_AEC_SU,hong2021egu_net,li2021selfSuperv_deepNMF_SU}.
An approach to learn learn dynamical models for the spectra of individual materials has also been proposed~\cite{drumetz2020learningEndmemberDynamics}.
Finally, other methods have been proposed using, e.g., Wasserstein~\cite{min2021metricLearningNet_SU}, adversarial~\cite{jin2021adversarialAEC_SU}, or cycle-consistency~\cite{gao2021cycu_neT_SU} loss functions during training. Deep priors~\cite{rasti2021HU_deepImagePrior} and unfolding optimization-based neural networks have also been considered~\cite{zhou2021ADMM_SU_networks}.

We highlight, however, that despite these advances previous MTHU methods did not address spatial EM variability without supervision, and also did not exploit deep learning frameworks. In the following, we will present a probabilistic model representing both spatial and temporal EM variability, and develop an unsupervised inference strategy based on RNNs.

\section{Overview of the proposed approach}
\label{sec:methodOverview}

We consider a probabilistic framework for MTHU. This amounts to two steps. The first is the \emph{modeling step}, which \cblue{consists in} defining a set of probability density functions (PDFs) which describe how the abundances and EMs evolve over time, and how the pixels are generated. Note that the model PDFs typically depend on different deterministic parameters, some of which are specified a priori (which we refer to as {\em hyperparameters}) and some which we intend to learn from the observed HIs. We represent the parameters to be learned in the set $\theta$. We include the subscript $\theta$ on the model PDFs in order to make their dependence on these parameters explicit.

The second is the \emph{inference step}, which consists in computing (an approximation of) the posterior distribution, which is the PDF of the abundances and EMs conditioned on the observed pixels. The inference step is also decomposed in two distinct parts which are interdependent. The first part consists in computing the approximate posterior distribution, while the second part involves learning the deterministic parameters of the model in $\theta$ by maximizing the likelihood of the pixels.

Note that this process leads to an \emph{unsupervised learning} problem, that is, the model parameters in $\theta$ and posterior distribution are both computed based only on the observed HI pixels (i.e., there is no separate training and testing data). In the following, we provide a high-level description of the approach, which is illustrated in Figures~\ref{fig:overall_diagram} and~\ref{fig:graphical_models}. Modeling and inference steps are then detailed in Sections~\ref{sec:mymodel} and~\ref{sec:proposedMethod}.

\paragraph*{\textbf{Modeling step}}
The first step is to characterize the dynamical evolution of the EMs and abundances. Under a Markovity assumption, it can be expressed using the following sequence of conditional probability distributions~\cite{sarkka2013bayesian}:
\begin{align}
    (\ba_{t},\bM_{t}) & \sim p_{\theta}(\ba_{t},\bM_{t}|\ba_{t-1},\bM_{t-1}) \,,
    \label{eq:pdf_dynamical}
\end{align}
where $\ba_t=[\ba_{1,t}^\top,\ldots,\ba_{N,t}^\top]^\top$ denotes the abundance maps in lexicographic ordering and $\bM_{t}=\big\{\bM_{1,1},\ldots,\bM_{N,t}\big\}$ the collection of EM matrices for all pixels, and $(\ba_{0},\bM_{0})\sim p_{\theta}(\ba_{0},\bM_{0})$. 
We also assume that the abundances and endmembers at time $t$ are statistically independent when conditioned on the their values at time $t-1$, that is,
\begin{align}\label{eq:aM_cond_ind}
    p_{\theta}(\ba_{t},\bM_{t}|\ba_{t-1},\bM_{t-1})
    = p_{\theta}(\ba_{t}|\ba_{t-1}) p_{\theta}(\bM_{t}|\bM_{t-1}) \,.
\end{align}
This allows us to model the evolution of $\ba_{t}$ and $\bM_{t}$ separately.

The second part of the model represents how the HI pixels are generated from $\ba_{t}$ and $\bM_{t}$, which is given by
\begin{align}
    \by_t & \sim p_{\theta}(\by_t|\ba_{t},\bM_{t}) \,,
    \label{eq:pdf_measurment}
\end{align}
where $\by_t=[\by_{1,t}^\top,\ldots,\by_{N,t}^\top]^\top$ denotes the HI in lexicographic ordering. Note that the pixels $\by_t$ are assumed to be conditionally independent given $\ba_{t}$ and $\bM_{t}$. These PDFs are defined explicitly in Section~\ref{sec:mymodel}.

\paragraph*{\textbf{Inference step}}
This step, which constitutes the solution to the MTHU problem, consists in computing the posterior PDF of the EMs and of the abundances given the observed pixels, which is given by:
\begin{align}
    & p_{\theta}(\cblue{\ba_{1},\bM_1},\ldots,\ba_T,\bM_{T}|\by_1,\ldots,\by_T) 
    \label{eq:posterior} \\
    & = \frac{p_{\theta}(\ba_0,\bM_0) \prod_{t=1}^T p_{\theta}(\by_t|\ba_{t},\bM_{t}) p_{\theta}(\ba_{t}|\ba_{t-1}) p_{\theta}(\bM_{t}|\bM_{t-1})}{p_{\theta}(\by_1,\ldots,\by_T)} \,, \nonumber
\end{align}
where the r.h.s. of~\eqref{eq:posterior} was obtained using the Bayes rule and the factorization in~\eqref{eq:pdf_dynamical}--\eqref{eq:pdf_measurment}. \cblue{The PDF in~\eqref{eq:posterior} generally does not have a closed form solution~\cite{sarkka2013bayesian}.} %
\cblue{One efficient solution is to use variational inference based on SGD~\cite{kingma14VAEs}, which attempts} to find an approximate posterior $q\in\calQ$ within a family of distributions $\calQ$ that is as close as possible to the true posterior in~\eqref{eq:posterior}. This approximation is often obtained by maximizing a lower bound on the marginal log-likelihood, the so-called {\em evidence lower bound} (ELBO):
\begin{align}
    \label{eq:first_elbo_illustrative}
    {\rm ELBO}(q,\theta) \leq \log \, p_{\theta}(\by_1,\ldots,\by_T) \,,
\end{align}
see Section~\ref{sec:proposedMethod} for a detailed explanation.
Under specific conditions over the model and posterior family $\calQ$, such as assuming conditionally Gaussian distributions, the maximization $\max_{q\in\calQ} {\rm ELBO}(q,\theta)$ can be solved locally using SGD techniques, which are computationally efficient when compared to solutions based on Monte Carlo sampling~\cite{kingma14VAEs}.

The flexibility provided by the family of posterior distributions $\calQ$ is paramount for the performance of the strategy.
Recent works have considered neural networks, parameterized by $\phi$, to represent the approximate posterior. Thus, instead of searching for $q$ in a (continuous) family of distributions $\calQ$, we search for the parameters $\phi$, such that the parameterized posterior, denoted by $q_{\phi}$, maximizes the ELBO.
Thus, the optimization becomes $\max_{\phi} {\rm ELBO}(q_{\phi},\theta)$.
For problems with a temporal Markov structure, RNNs provide a parameterization that, although flexible, is computationally efficient. Moreover, they explicitly explore the temporal structure in the data, having shown excellent performance in various sequence modeling tasks~\cite{lipton2015criticalReviewRNNs}. This will motivate us to use an RNN in the parametrization of our posterior in Section~\ref{ssec:rnn_implementation}.

Finally, the parameters of the generative model, $\theta$, are also learned within the same framework. The underlying idea is to \cblue{perform type-II maximum likelihood (ML) estimation~\cite{murphy2012machineLearningBook},} that is, $\max_{\theta} \log p_{\theta}(\by_1,\ldots,\by_T)$.
However, since computing $p_{\theta}(\by_1,\ldots,\by_T)$ is intractable, $\theta$ is also computed by maximizing the ELBO w.r.t. $\theta$ using SGD. Thus, as ${\rm ELBO}(q_{\phi},\theta)$ is maximized w.r.t. $q_{\phi}$, the lower bound in~\eqref{eq:first_elbo_illustrative} becomes tighter and its maximization w.r.t. $\theta$ better approximates ML estimation. Thus, the complete inference problem is formulated as the maximization of the ELBO w.r.t. both the posterior and the model parameters, that is, $\max_{\theta, \, \phi} {\rm ELBO}(q_{\phi},\theta)$.

\section{Proposed model}
\label{sec:mymodel}

The modeling step will be divided as follows.
First, we develop a mixing model and represent EM variability (which defines $p_{\theta}(\by_t|\ba_{t},\bM_{t})$). Next, we consider the dynamical behavior of the EMs, and finally of the abundances (which define $p_{\theta}(\bM_{t}|\bM_{t-1})$ and $p_{\theta}(\ba_{t}|\ba_{t-1})$, respectively). 

\subsection{Mixture model with EM variability}

As discussed in Section~\ref{sec:relatedw}, devising EM models that combine flexibility to represent complex spectral variability with simplicity of having a small number of parameters is challenging.
Flexible models such as the PLMM~\cite{Thouvenin_IEEE_TSP_2016_PLMM}, GLMM~\cite{imbiriba2018GLMM} and ALMM~\cite{hong2019augmentedLMMvariability} have many degrees of freedom and require additional regularization strategies to guarantee physically meaningful solutions, while simpler models such as the ELMM~\cite{drumetz2016blindUnmixingELMM} are too restrictive to represent complex spectral variability.
An important information which can be used in the design of mixing models accounting for endmember variability is the spectral correlation of EM signatures~\cite{borsoi2020variabilityReview2}. This points to a natural representation of EMs as smooth functions. Although the smoothness of the EMs can be introduced through regularization (see, e.g., \cite{Borsoi_2018_Fusion,halimi2016unmixingVariabilityNonlinearityMismodeling}), this leads to high-dimensional and potentially costly HU solutions. On the other hand, a more efficient and interpretable model can be obtained by directly parametrizing smoothness using properly selected basis functions~\cite{Halimi_IEEE_Trans_CI_2017}.

In this work, we consider an EM model inspired by the GLMM~\cite{imbiriba2018GLMM}, which represents spectral variability using a multiplicative scaling of reference EM spectra that vary for each band, endmember and pixel. However, instead of using regularizations we propose to constrain the scaling factors to be linear combinations of spectrally smooth functions. The resulting model, which we call Smooth GLMM (SGLMM), represents each observed HI pixel $\by_{n,t}$ as follows:
\begin{equation} \label{eq:model_lrglmm}
    \by_{n,t} = 
    \underbrace{\big(\bM_0\odot(\mathbb{1}+\bD \widetilde{\bPsi}_{n,t})\big)}_{\bM_{n,t}}
     \ba_{n,t} +
    \br_{n,t} \,,
\end{equation}
where $\odot$ represents the Hadamard (elementwise) product, $\mathbb{1}$ is an $L\times P$ matrix of ones, $\bM_0\in\amsmathbb{R}^{L\times P}$ a set of reference or average EM signatures, matrix $\bD\in\amsmathbb{R}^{L\times K}$ contains $K$ spectrally smooth basis vectors as its columns, and $\widetilde{\bPsi}_{n,t}\in\amsmathbb{R}^{K\times P}$ contains the low-dimensional coefficients that parameterize the variability of each EM. Vector $\br_{n,t}\in\amsmathbb{R}^L$ denotes zero-mean additive Gaussian noise.

It is instructive to analyze how the variability of the endmembers $\bM_{n,t}$ is introduced in the model~\eqref{eq:model_lrglmm}. %
The EM matrix $\bM_{n,t}$ is formed by scaling the reference EMs in $\bM_0$ bandwise by matrix $\mathbb{1}+\bD\widetilde{\bPsi}_{n,t}\in\amsmathbb{R}^{L\times P}$. This model is similar to the GLMM, the difference being in the structure of this multiplicative scaling matrix. First, note that when $\widetilde{\bPsi}_{n,t}\approx\cb{0}$, the term $\bD\widetilde{\bPsi}_{n,t}$ is also small and the scaling factors will be close to $\mathbb{1}$, meaning that $\bM_{n,t}\approx\bM_0$, i.e., the spectral variability is small. Thus, the amount of EM variability depends directly on the amplitude of the elements of $\widetilde{\bPsi}_{n,t}$. Second, matrix $\bD\widetilde{\bPsi}_{n,t}$ represents a perturbation over the constant scaling $\mathbb{1}$, and its properties depend directly on the choice of $\bD$. Thus, by
properly selecting $\bD$ we can constrain $\bD\widetilde{\bPsi}_{n,t}$ to represent smooth functions with few parameters, leading to smooth spectral variations in $\bM_{n,t}$. Following an idea used in~\cite{Halimi_IEEE_Trans_CI_2017} for robust HU with smooth additive residual terms, we select the columns of $\bD$ as the first $K$ rows of the discrete cosine transform (DCT) matrix. 

The SGLMM models endmember variability using $KP$ parameters. The number of basis functions $K$ gives a trade-off between existing models in the literature: when $K=1$, $\bD$ will contain only a constant vector and the model becomes equivalent to the ELMM~\cite{drumetz2016blindUnmixingELMM}, whereas for $K=L$ it has the same flexibility as the PLMM~\cite{Thouvenin_IEEE_TSP_2016_PLMM} and GLMM~\cite{imbiriba2018GLMM}. Values of $K\ll L$ should give the SGLMM sufficient flexibility to represent smooth spectral variability accurately.

Representing the spectral variability parameters in vectorized form as $\bpsi_{n,t}=\vect(\widetilde{\bPsi}_{n,t})$ and assuming the noise $\br_{n,t}$ to be independent for each pixel, the PDFs in the generative model~\eqref{eq:pdf_measurment} can be rewritten equivalently in terms of $\bpsi_{n,t}$ as 
\begin{align}
    & p_{\theta}(\by_{t}| \ba_{t},\bpsi_{t}) = \prod_{n=1}^N p_{\theta}(\by_{n,t}| \ba_{n,t},\bpsi_{n,t}) \,,
    \label{eq:meas_model_iid}
\end{align}
where
\begin{align}
    & p_{\theta}(\by_{n,t}| \ba_{n,t},\bpsi_{n,t}) 
    \nonumber\\ 
    & \hspace{0.6cm}
    = \calN\Big(\big(\bM_0\odot(\mathbb{1}+\bD\vect^{-1}(\bpsi_{n,t}))\big)\ba_{n,t}, \,\sigma_r^2\bI\Big) \,,
    \label{eq:meas_model_pixelwise}
\end{align}
\cblue{in which $\sigma_r\in\amsmathbb{R}_+^*$} is the standard deviation of the measurement noise, which is assumed to be independent and identically distributed for different bands.

\subsection{Dynamical model for the EM scaling parameters}

In this work, we consider $\bM_0$ to be a deterministic parameter of the model and estimated from the observed HIs using an approximate ML framework (i.e., $\bM_0\in\theta$). This means that the EM matrix for each $t$ and $n$, $\bM_{n,t}=\bM_0\odot(\mathbb{1}+\bD \widetilde{\bPsi}_{n,t})$, is
a deterministic function of the lower-dimensional vector of scaling factors $\bpsi_{n,t}$. Thus, we can substitute the problem of estimating the very high-dimensional $p_{\theta}(\bM_{t}|\bM_{t-1})$ by the problem of estimating $p_{\theta}(\bpsi_t|\bpsi_{t-1})$, with $\bpsi_t=\big[\bpsi_{1,t}^\top,\ldots,\bpsi_{N,t}^\top\big]^\top$. 
Since vector $\bpsi_t$ is still high dimensional, we consider an independence assumption on the time evolution between different pixels:
\begin{align}
    p_{\theta}(\bpsi_t|\bpsi_{t-1}) = \prod_{n=1}^N p_{\theta}(\bpsi_{n,t}|\bpsi_{n,t-1}) \,,
    \label{eq:dynamical_mdl_psi_iid_px}
\end{align}
for $t\geq1$, where the prior PDF for time instant $t=0$ will be specified later in Section~\ref{ssec:complete_model}. We consider a Gaussian distribution to represent the evolution of $\bpsi_{n,t}$:
\begin{align}
    p_{\theta}(\bpsi_{n,t}|\bpsi_{n,t-1}) = \calN\big(\bpsi_{n,t-1}, \sigma_{\psi}^2\bI_{PK}\big) \,,
    \label{eq:dynamical_mdl_psi}
\end{align}
\cblue{where $\sigma_{\psi}\in\amsmathbb{R}_+^*$} is the distribution standard deviation, which controls its uncertainty and is assumed to be isotropic.
Note that the evolution of $\bpsi_{n,t}$ is not assumed to be affected by abrupt changes, which leads us to consider $\sigma_{\psi}$ constant. This assumption is motivated from the fact that the reflectance of materials are primarily influenced by their physico-chemical composition (e.g., particle size and roughness in packed particle spectra~\cite{Hapke1981}, or biophysical parameters in leaf spectra~\cite{jacquemoud2001leafOpticalPropertiesReview}), which we assume to change smoothly at fine time scales.

\vspace{-0.3cm}
\subsection{Abundances model}

In order to represent the abundances dynamical behavior, we first assume their time evolution to be independent for different pixels in order to make the problem tractable, that is,
\begin{align}
    p_{\theta}(\ba_{t}|\ba_{t-1}) = \prod_{n=1}^N p_{\theta}(\ba_{n,t}|\ba_{n,t-1}) \,,
    \label{eq:dynamical_mdl_a_iid_px}
\end{align}
for $t\geq1$, where the prior PDF for time instant $t=0$ will be specified later in Section~\ref{ssec:complete_model}.
To represent the time evolution at each pixel, we consider a Dirichlet distribution. The Dirichlet is a natural choice of distribution to model the abundances as it enforces the physical constraints that the elements of $\ba_t$ should be nonnegativity and sum to one~\cite{Halimi_IEEE_TIP_2015,eches2012adaptiveMRFunmixing}. The transition PDF is then given by
\begin{align}
    \label{eq:abundances_time_evol}
    p_{\theta}(\ba_{n,t}|\ba_{n,t-1}) = \Dir(\balpha_{n,t}) \,,
\end{align}
where $\balpha_{n,t}\in\amsmathbb{R}^{P}_+$ denotes the concentration parameters, which are a function of the abundances at the previous time instant, $\ba_{n,t-1}$ (i.e., the parameters of $p_{\theta}(\ba_{n,t}|\ba_{n,t-1})$, $\balpha_{n,t}$, are a function of the conditioning variable).
Note that the uncertainty of the abundances predictions is represented implicitly in $\balpha_{n,t}$, where small concentration values yield low uncertainty and temporally smooth transition, whereas large values lead to higher uncertainty, allowing for more changes.

However, the Dirichlet distribution can make inference difficult. One workaround consists of using, e.g., Laplace's method, which approximates the Dirichlet distribution 
$\Dir(\balpha_{n,t})$
by a Gaussian with mean and inverse covariance equal to the mode of the original distribution and the Hessian of its negative logarithm, respectively~\cite{barber2012ML_book}. However, since the Dirichlet distribution is supported at the simplex, this approximation can be inaccurate. To overcome this problem, MacKay~\cite{mackay1998choiceBasisDirichlet} proposed to perform this approximation in the so-called \textit{softmax basis}, which consists in a mapping $\bpi^{-1}:\ba_{n,t}\mapsto\bc_{n,t}$ from the unity simplex to $\amsmathbb{R}^P$, where $\bpi$ is the softmax function:
\begin{align}
    \bpi^{-1}(\ba_{n,t}) & = \bc_{n,t}\,,
    \\
    \pi_i(\bc_{n,t}) & = \frac{\exp(c_{n,t,i})}{\sum_j \exp(c_{n,t,j})} \,, \,\,\, i\in\{1,\ldots,P\} \,,
\end{align}
with $\pi_i$, $c_{n,t,i}$ being the $i$-th positions of $\bpi$ and $\bc_{n,t}$. This approximation is very accurate and has been used in several works to facilitate statistical inference~\cite{hennig2012kernelTopicModels,srivastava2017autoencodingTopicModels}.
Thus, replacing $\ba_{n,t}$ by the softmax parameters $\bc_{n,t}$, we achieve the following alternative representation of~\eqref{eq:abundances_time_evol}~\cite{mackay1998choiceBasisDirichlet}:
\begin{align}
    p_{\theta}(\bc_{n,t}| & \bc_{n,t-1}) 
    \nonumber \\
    ={}& \frac{\Gamma(\sum_{i=1}^{P}\alpha_{n,t,i})}{\prod_{i=1}^P\Gamma(\alpha_{n,t,i})} \prod_{i=1}^P \pi_i(\bc_{n,t})^{\alpha_{n,t,i}} g(\cb{1}^\top\bc_{n,t}) \,,
    \label{eq:dirichlet_softmax_interm1}
\end{align}
where $\alpha_{n,t,i}$ is the $i$-th position of $\balpha_{n,t}$, $\Gamma$ denotes the Gamma function, and $g$ is an arbitrary distribution used to constrain an extra degree of freedom (since the Dirichlet has only $P-1$ degrees of freedom), selected as $g(x)\propto\exp(-\frac{\epsilon}{2}x^2)$ for mathematical convenience~\cite{mackay1998choiceBasisDirichlet}. 
The Gaussian approximation of this distribution is then given by $p_{\theta}(\bc_{n,t}|\bc_{n,t-1}) \approx \calN(\bmu_{n,t},\bSigma_{n,t})$~\cite{mackay1998choiceBasisDirichlet,hennig2012kernelTopicModels}, with $\bmu_{n,t}$ given by
\begin{align}
    \mu_{n,t,i} ={} & \log(\alpha_{n,t,i}) -\frac{1}{P} \sum_{\ell=1}^P \log(\alpha_{n,t,\ell}) \,,
\end{align}
where $\mu_{n,t,i}$ is the $i$-th position of $\bmu_{n,t}$ and $\bSigma_{n,t}$ is the negative Hessian of~\eqref{eq:dirichlet_softmax_interm1} at $\bc_{n,t}=\bmu_{n,t}$.
Note that the mean and covariance $\bmu_{n,t}$ and $\bSigma_{n,t}$ are a function of $\balpha_{n,t}$ and, consequently, depend implicitly on $\bc_{n,t-1}$.

Therefore, we can approximate the transition PDF~\eqref{eq:abundances_time_evol} by a Gaussian one on the softmax basis. Note that the relationship between the parameters of both models, that is, between $\balpha_{n,t}$ and $\bmu_{n,t}$ and $\bSigma_{n,t}$, is nonlinear and burdensome to compute. However, by working on the softmax basis we do not need to specify $p_{\theta}(\ba_{n,t}|\ba_{n,t-1})$ explicitly in \eqref{eq:abundances_time_evol}.
Instead, we can directly define the Gaussian transition for $p_{\theta}(\bc_{n,t}|\bc_{n,t-1})$, which is mathematically more convenient. This implicitly defines a transition probability $p_{\theta}(\ba_{n,t}|\ba_{n,t-1})$ by mapping $\bc_{n,t}\sim p_{\theta}(\bc_{n,t}|\bc_{n,t-1})$ into the simplex, approximating a Dirichlet distribution. Thus, we consider the following model:
\begin{align}
    p_{\theta}(\bc_{n,t}|\bc_{n,t-1}) = \calN\big(\bc_{n,t-1}, \, \sigma_a^2(\bc_{n,t-1})\bI_P\big) \,,
    \label{eq:abundances_time_evol_softmax}
\end{align}
where $\bI_P$ is a $P\times P$ identity matrix.
Note that, for simplicity, the covariance matrix in~\eqref{eq:abundances_time_evol_softmax} was constrained to be isotropic, and is scaled by $\sigma_a^2(\bc_{n,t-1})$. The \cblue{function $\sigma_a:\amsmathbb{R}^P\to\amsmathbb{R}_+^*$} computes the standard deviation of each element of $\bc_{n,t}\sim p_{\theta}(\bc_{n,t}|\bc_{n,t-1})$ as a function of $\bc_{n,t-1}$. Thus, it directly influences the amount of change in the abundances: the larger $\sigma_a(\bc_{n,t-1})$, the larger the changes we expect to observe between $\bc_{n,t-1}$ and $\bc_{n,t}$. This function, which is part of the generative model, will be learned during inference using a maximum likelihood approach. It will be parameterized using a fully connected neural network with $R_{\sigma_a}$ layers, where each hidden layer has $P$ neurons and uses the ReLU activation function, and the output layer maps to a scalar and uses an exponential activation function to ensure the output is positive.

\subsection{The complete model}
\label{ssec:complete_model}

To finish the model derivation, we need to define the initial PDFs at time $t=0$, which under the new parametrization of the abundances and endmembers, which we assume to be pixelwise independent Gaussian distributions, given by:
\begin{align}
    p_{\theta}(\bc_{0},\bpsi_{0}) &= \prod_{n=1}^N p_{\theta}(\bc_{n,0})p_{\theta}(\bpsi_{n,0}) \,,
    \label{eq:final_mdl_init_iid} 
    \\
    p_{\theta}(\bc_{n,0}) &= \calN\big(\bnu_{0}^c,\diag(\bgamma_{0}^c)^2\big) \,,
    \label{eq:final_mdl_init_a}
    \\
    p_{\theta}(\bpsi_{n,0}) &= \calN\big(\bnu_{0}^{\psi},\diag(\bgamma_{0}^\psi)^2\big) \,,
    \label{eq:final_mdl_init_psi}
\end{align}
for all $n=1,\ldots,N$, where the means $\bnu_{0}^c$, $\bnu_{0}^\psi$ and diagonal covariance parameters, $\bgamma_{0}^c$, $\bgamma_{0}^\psi$ are constant and shared among all pixels in order to reduce the amount of parameters in the model. Finally, the measurement model~\eqref{eq:meas_model_pixelwise} can be written using the softmax abundance reparametrization as:
\begin{align}
    & p_{\theta}(\by_{n,t}| \bc_{n,t},\bpsi_{n,t}) 
    \nonumber\\ & %
    = \calN\Big(\big(\bM_0\odot(\mathbb{1}+\bD\vect^{-1}(\bpsi_{n,t}))\big)\bpi(\bc_{n,t}), \,\sigma_r^2\bI\Big) \,,
    \label{eq:meas_model_pixelwise_softmax}
\end{align}
where $\bpi$ is the softmax function.

\begin{figure}[t]
    \centering
    \includegraphics[width=0.7\linewidth]{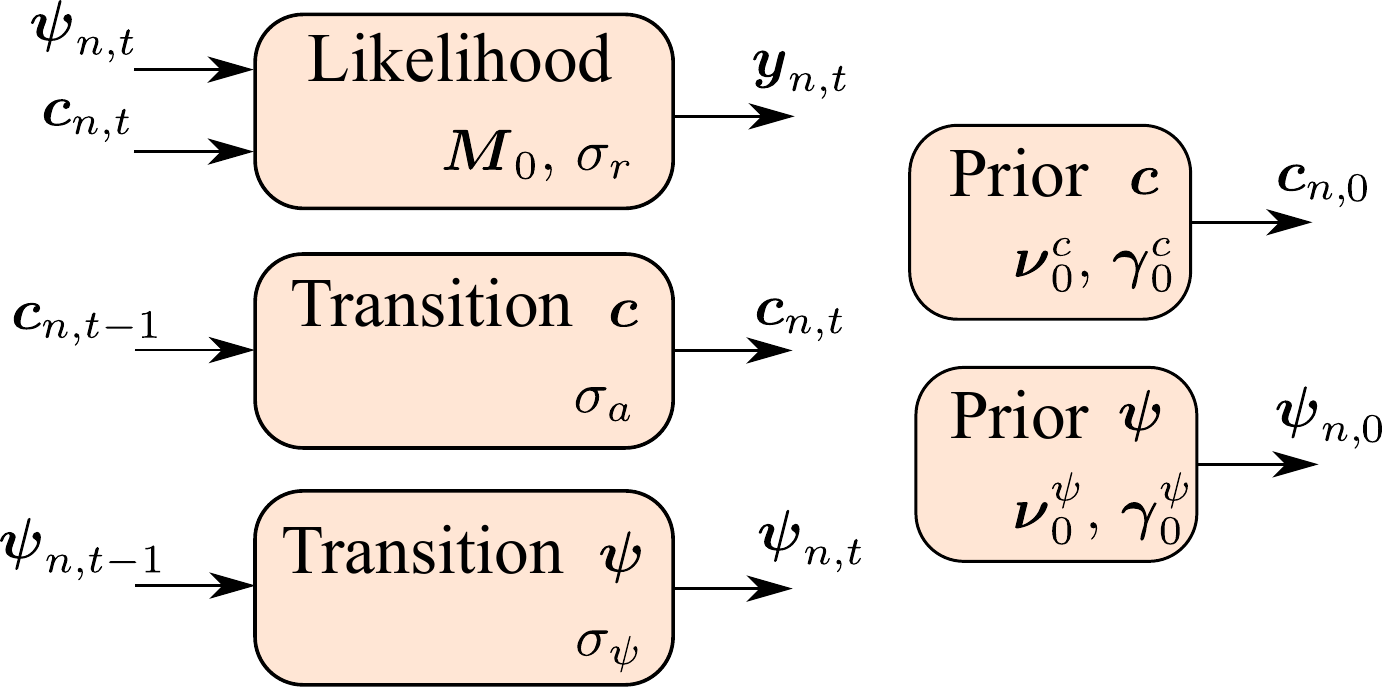}
    \vspace{-1.5ex}
    \caption{Illustrative diagram of the proposed generative model.}
    \label{fig:illustrative_generative}
\end{figure}

The final dynamical model is then given by equations~\eqref{eq:final_mdl_init_a},~\eqref{eq:final_mdl_init_psi} (initial PDFs),~\eqref{eq:abundances_time_evol_softmax},~\eqref{eq:dynamical_mdl_psi} (the dynamical model) and~\eqref{eq:meas_model_pixelwise_softmax} (the measurement model).
Finally, we denote the parameters of the model which will be estimated from the data using approximate ML inference by $\theta=\{\bM_0,\sigma_r,\sigma_a,\bnu_0^c,\bnu_0^\psi,\bgamma_0^c,\bgamma_0^\psi\}$. An illustrative diagram of the proposed generative model can be seen in Figure~\ref{fig:illustrative_generative}.

\section{Variational inference with RNNs for HU}
\label{sec:proposedMethod}

In this section, we will present the proposed solution to the inference step, referred to as ReDSUNN. %
Considering the parametrization of the abundances and of endmember variability derived in the previous section, this task, which consists in performing MTHU, becomes that of approximating the posterior distribution:
\begin{align}
    p_{\theta}\big(\cblue{\bc_{1},\bpsi_1,}\ldots,\bc_T,\bpsi_{T} \big| \by_1,\ldots,\by_T\big) \,,
    \label{eq:posterior_new}
\end{align}
where $\bc_t=\big[\bc_{1,t}^\top,\ldots,\bc_{N,t}^\top\big]^\top$.
First, let us denote with an underline the collection of variables at all time instants:
\begin{align}
    \underline\by &= \{\by_1,\ldots,\by_T\} \,,
    \\
    \underline\by_n &= \{\by_{n,1},\ldots,\by_{n,T}\} \,,
    \\
    \underline\bc &= \{\bc_0,\ldots,\bc_T\} \,,
    \\
    \underline\bpsi &= \{\bpsi_0,\ldots,\bpsi_T\} \,.
\end{align}
As discussed in Section~\ref{sec:methodOverview}, due to the nonlinearity in the model caused by the interaction between the abundances and the variability scaling factors, and the potentially high dimensionality of these variables, it is not possible to compute the posterior distribution~\eqref{eq:posterior_new} in closed form. In this work, we adopt a deep variational inference framework: we consider a parametric surrogate distribution $q_{\phi}(\underline\bc,\underline\bpsi|\underline\by)$ from a sufficiently flexible family with parameters $\phi$, and learn its parameters by minimizing the Kullback-Leibler (KL) divergence between $q_{\phi}(\underline\bc,\underline\bpsi|\underline\by)$ and the true posterior $p_{\theta}(\underline\bc,\underline\bpsi|\underline\by)$:
\begin{align}
    \KL\big(p_{\theta}(\underline\bc,\underline\bpsi|\underline\by) \big\| & q_{\phi}(\underline\bc,\underline\bpsi|\underline\by)\big) = \log p_{\theta}(\underline\by)
    \nonumber \\
    & + \Ex_{q_{\phi}(\underline\bc,\underline\bpsi|\underline\by)}\{\log q_{\phi}(\underline\bc,\underline\bpsi|\underline\by)\}
    \nonumber \\
    & - \Ex_{q_{\phi}(\underline\bc,\underline\bpsi|\underline\by)}\{\log p_{\theta}(\underline\bc,\underline\bpsi,\underline\by)\} \,.
\end{align}
Since the KL divergence is nonnegative and $\log p_{\theta}(\underline\by)$ is a constant, the above expression can be equivalently minimized by maximizing a lower bound to the data likelihood formed by the last two terms in the right hand side of the expression, which is the so-called ELBO~\cite{barber2012ML_book,kingma2014VAEs}:
\begin{align}
    \log p_{\theta}(\underline\by) \geq{} & \Ex_{q_{\phi}(\underline\bc,\underline\bpsi|\underline\by)}\{\log p_{\theta}(\underline\bc,\underline\bpsi,\underline\by)\}
    \nonumber \\
    & - \Ex_{q_{\phi}(\underline\bc,\underline\bpsi|\underline\by)}\{\log q_{\phi}(\underline\bc,\underline\bpsi|\underline\by)\} \,.
    \label{eq:ELBO_general}
\end{align}
Recent advances in variational deep learning such as in variational autoencoders has made it possible to devise efficient algorithms to maximize~\eqref{eq:ELBO_general} when the PDFs are possibly parameterized by deep neural network using, e.g., stochastic backpropagation algorithms~\cite{kingma2014adam}.
Furthermore, the conditional independence assumptions of the model in the previous section can be exploited to further simplify the inference problem. In the following, we factorize the ELBO both in the temporal as well as in the pixel dimensions.

Note that, as discussed in Section~\ref{sec:methodOverview},~\eqref{eq:ELBO_general} will be optimized both with respect to the parameters of the posterior distribution $\phi$, but also with respect to the parameters of the generative model $p_{\theta}(\underline\bc,\underline\bpsi,\underline\by)$ in the set $\theta$ in order to estimate them by approximate ML inference.
In the following, we will use the Markov and pixelwise conditional independence assumptions of the generative model in Section~\ref{sec:mymodel} to factorize $q_{\phi}(\underline\bc,\underline\bpsi|\underline\by)$ and simplify the solution to the inference problem.

\subsection{Factorizing the posterior distribution}

\paragraph*{\textbf{Factorizing the posterior distribution in time}}
Various kinds of parametrizations of the distribution $q_{\phi}(\underline\bc,\underline\bpsi|\underline\by)$ have been proposed. One of the simplest is to consider a mean field assumption~\cite{krishnan2017structuredInferenceNets}, which assumes that $\{\bc_t,\bpsi_t\}$ and $\{\bc_{t'},\bpsi_{t'}\}$ are conditionally independent given $\underline\by$, for all $t\neq t'$. However, this disregards the temporal structure of the data. A more suitable factorization can be obtained by noting that the Markov property of the model can be used to show that the true posterior factorizes as
\begin{align}
    p_{\theta}(\underline\bc,\underline\bpsi|\underline\by) = p_{\theta}(\bc_0,\bpsi_0|\underline\by) \prod_{t=1}^T p_{\theta}(\bc_t,\bpsi_t|\bc_{t-1},\bpsi_{t-1},\underline\by) \,.
    \nonumber
\end{align}
Incorporating this assumption into the variational approximation $q_{\phi}(\underline\ba,\underline\bpsi|\underline\by)$ leads to a similar factorization:
\begin{align}
    q_{\phi}(\underline\bc,\underline\bpsi|\underline\by) = q_{\phi}(\bc_0,\bpsi_0|\underline\by) \prod_{t=1}^T q_{\phi}(\bc_t,\bpsi_t|\bc_{t-1},\bpsi_{t-1},\underline\by) \,,
    \label{eq:posterior_fact_time}
\end{align}
which preserves the temporal dependency of the model. %

\paragraph*{\textbf{Factorizing the posterior distribution in pixels}}

The vectors $\bc_t,\bpsi_t$ in~\eqref{eq:posterior_fact_time} contain the abundances and variability coefficients for all image pixels, and are thus of very high dimension. Therefore, additional simplifications are necessary in order to make inference tractable. One important property is that in the model derived in Section~\ref{sec:mymodel}, the initial, transition, and measurement PDFs (equations~\eqref{eq:final_mdl_init_a},~\eqref{eq:final_mdl_init_psi},~\eqref{eq:abundances_time_evol_softmax},~\eqref{eq:dynamical_mdl_psi} and~\eqref{eq:meas_model_pixelwise_softmax}) can be factorized among the different image pixels. Thus, the inference process can be factorized at the pixel level, which leads to the following form for the posterior distribution:
\begin{align}
    & q_{\phi}(\bc_t,\bpsi_t|\bc_{t-1},\bpsi_{t-1},\underline\by)
    \nonumber \\
    & \qquad = \prod_{n=1}^N q_{\phi}(\bc_{n,t},\bpsi_{n,t}|\bc_{n,t-1},\bpsi_{n,t-1},\underline\by_{n}) \,,
    \label{eq:patchwise_indep_posterior}
\end{align}
for $t\geq 1$,
and similarly for the initial PDF at $t=0$:
\begin{align}
    q_{\phi}(\bc_0,\bpsi_0|\underline\by) 
    = \prod_{n=1}^N q_{\phi}(\bc_{n,0},\bpsi_{n,0}|\underline\by_{n}) \,.
    \label{eq:patchwise_indep_posterior_t0}
\end{align}
Although this factorization does not directly consider spatial correlation between different pixels, which has been found to be a useful source of prior information in HU~\cite{eches2011bayesianSpatialMarkovUnmixing,Halimi_IEEE_TIP_2015}, it allows us to work with pixelwise variational posterior PDFs (i.e., the r.h.s. of~\eqref{eq:patchwise_indep_posterior} and~\eqref{eq:patchwise_indep_posterior_t0}) which have a much lower dimension, thus reducing the computational burden associated with the inference step. To some extent, spatial information can still be introduced indirectly by constraining the parametrization of the variational posterior distributions among different pixels, which will be explained in the rest of this subsection. The incorporation of spatial information directly through the probabilistic model will be investigated in a future work.

\paragraph*{\textbf{Parameterizing the posterior}}

A key aspect of the model is how to parameterize the posterior PDFs of the different pixels in the r.h.s. of~\eqref{eq:patchwise_indep_posterior} and~\eqref{eq:patchwise_indep_posterior_t0}. First, variational inference implies selecting a parametric family of distributions from which to select $q_{\phi}$, which directly impact the results. Note that the true posterior in~\eqref{eq:posterior} might have a complex form and be possibly multimodal, however, its form is not known in advance. Thus, as in recent works in deep variational inference (see, e.g.,~\cite{kingma2014VAEs}) we considered a Gaussian family for $q_{\phi}$ since this will simplify the maximization of the ELBO considerably (through, e.g., the reparametrization trick and closed form expressions for KL divergences), leading to important computation savings. Thus, it can be expressed as:
\begin{align}
    & q_{\phi}\big(\bc_{n,t},\bpsi_{n,t}\big|\bc_{n,t-1},\bpsi_{n,t-1},\underline\by_{n}\big)
    \nonumber \\
    & \hspace{0.8cm} = 
    \calN\big(\bmu_{\phi}^{c,\psi}(\bUpsilon_{n,t}),\diag(\bsigma_{\phi}^{c,\psi}(\bUpsilon_{n,t}))^2\big) \,,
    \label{eq:posterior_fact2}
\end{align}
where $\bUpsilon_{n,t}=\big\{\bc_{n,t-1},\bpsi_{n,t-1},\underline\by_{n}\big\}$ and $\bmu_{\phi}^{c,\psi}$ and $\bsigma_{\phi}^{c,\psi}$ are functions (e.g., neural networks parameterized by $\phi$) which compute the parameters of the posterior distribution, mapping the data $\{\bc_{n,t-1},\bpsi_{n,t-1},\underline\by_{n}\}$ to the mean and the square root of the diagonal covariance matrix of the Gaussian posterior, respectively.  
For convenience of notation, we decompose $\bmu_{\phi}^{c,\psi}$ and $\bsigma_{\phi}^{c,\psi}$ into two functions:
\begin{align}
    \bmu_{\phi}^{c,\psi} = 
    \begin{bmatrix}
    \bmu_{\phi}^{c} \\
    \bmu_{\phi}^{\psi}
    \end{bmatrix} 
    \,, \qquad 
    \bsigma_{\phi}^{c,\psi} = 
    \begin{bmatrix}
    \bsigma_{\phi}^{c} \\
    \bsigma_{\phi}^{\psi}
    \end{bmatrix} \,.
    \label{eq:split_mean_cov_a_psi}
\end{align}
Note that functions $\bmu_{\phi}^{c}$ and $\bsigma_{\phi}^{c}$ compute the mean and square root of the diagonal covariance matrix of the $q_{\phi}\big(\bc_{n,t}\big|\bc_{n,t-1},\bpsi_{n,t-1},\underline\by_{n}\big)$, while $\bmu_{\phi}^{\psi}$ and $\bsigma_{\phi}^{\psi}$ compute the mean and square root of the diagonal covariance matrix of $q_{\phi}\big(\bpsi_{n,t}\big|\bc_{n,t-1},\bpsi_{n,t-1},\underline\by_{n}\big)$.

A Gaussian parametrization is also used for the posterior distribution of the initial PDF:
\begin{align}
    q_{\phi}(\bc_0,\bpsi_0|\underline\by)=\calN\big(\bzeta^{c,\psi},\diag(\bxi^{c,\psi})^2\big) \,,
    \label{eq:posterior_fact2_init}
\end{align}
where a fixed distribution was used for all pixels, with $\bzeta^{c,\psi}$ and $\bxi^{c,\psi}$ being the mean and the diagonal of the square root of the covariance matrix, respectively.

An important observation is that we consider a \emph{shared parametrization}, where the posterior in~\eqref{eq:posterior_fact2} and~\eqref{eq:posterior_fact2_init} has the same form for all pixels. More precisely, this means that the same functions $\bmu_{\phi}^{c,\bpsi}$ and $\bsigma_{\phi}^{c,\psi}$ are used to compute the posterior mean and covariance for every HI pixel, given the input data $\bUpsilon_{n,t}$. This is an important characteristic of the method, since it allows information from different pixels (i.e., from the whole image) to be leveraged jointly in the estimation of the model and, consequently, of the abundances and variability coefficients in each pixel, $\bc_{n,t},\bpsi_{n,t}$, $n=1,\ldots,N$.

\subsection{Factorizing the ELBO cost function}

Using the simplifications derived in the previous subsection, in the following we will rewrite the ELBO cost function~\eqref{eq:ELBO_general} in terms of the factorized model.

\paragraph*{\textbf{Factorizing the ELBO temporally}}

Using the factorization~\eqref{eq:posterior_fact_time} and the Markovity of the model, the lower bound in~\eqref{eq:ELBO_general} can be written as~\cite{krishnan2017structuredInferenceNets}:
\begin{align}
    & \log p_{\theta}(\underline\by) \geq \calL(\theta,\phi,\underline\by) 
    = \sum_{t=1}^T \Ex_{q_{\phi}(\bc_t,\bpsi_t|\underline\by)} \big\{\log p_{\theta}(\by_t|\bc_t,\bpsi_t)\big\}
    \nonumber \\ & 
    - \KL\big(q_{\phi}(\bc_0,\bpsi_0|\underline\by) \big\| p_{\theta}(\ba_0,\bpsi_0)\big)
    - \sum_{t=1}^T \Ex_{q_{\phi}(\bc_{t-1},\bpsi_{t-1}|\underline\by)} \Big\{
    \nonumber\\
    & \KL\big(q_{\phi}(\bc_t,\bpsi_t|\bc_{t-1},\bpsi_{t-1},\underline\by) \big\| p_{\theta}(\bc_t,\bpsi_t|\bc_{t-1},\bpsi_{t-1})\big) \Big\} \,.
    \label{eq:ELBO_factorized_time}
\end{align}

\paragraph*{\textbf{Factorizing at pixel level}}

Using the pixelwise factorization of the generative and posterior PDFs discussed in Section~\ref{sec:mymodel}, we can simplify each term of~\eqref{eq:ELBO_factorized_time}. To this end, we use the fact that $\KL(p(x_1,x_2)\|q(x_1,x_2))=\KL(p(x_1)\|q(x_1))+\KL(p(x_2)\|q(x_2))$ when both $p(x_1,x_2)=p(x_1)p(x_2)$ and $q(x_1,x_2)=q(x_1)q(x_2)$ are independent, and the fact that $\Ex_{p(x_1,x_2)}\{f(x_1)\}=\Ex_{p(x_1)}\{f(x_1)\}$. We proceed to analyse each term of~\eqref{eq:ELBO_factorized_time} in the following.

\textbf{First term:} Using the factorization of the measurement model in~\eqref{eq:meas_model_iid}, the first term in the r.h.s. of~\eqref{eq:ELBO_factorized_time} becomes:
\begin{align}
    & \Ex_{q_{\phi}(\bc_t,\bpsi_t|\underline\by)} \big\{\log p_{\theta}(\by_{t}|\bc_{t},\bpsi_{t})\big\}
    \nonumber\\ &
    = \sum_{n=1}^N \Ex_{q_{\phi}(\bc_{n,t},\bpsi_{n,t}|\underline\by_n)} \big\{\log p_{\theta}(\by_{n,t}|\bc_{n,t},\bpsi_{n,t})\big\} \,.
\end{align}

\textbf{Second term:} Using the pixelwise independence of the initial PDF in the generative model~\eqref{eq:final_mdl_init_iid} and in the posterior~\eqref{eq:patchwise_indep_posterior_t0}, the KL divergence can be written as:
\begin{align}
    \KL & \big(q_{\phi}(\bc_{0},\bpsi_{0}|\underline\by) \big\| p_{\theta}(\bc_{0},\bpsi_{0})\big)
    \nonumber \\ &
    = \sum_{n=1}^N \KL\big(q_{\phi}(\bc_{n,0},\bpsi_{n,0}|\underline\by_{n}) \big\| p_{\theta}(\bc_{n,0},\bpsi_{n,0})\big) \,.
\end{align}

\textbf{Third term:} Using the pixelwise independence of the posterior~\eqref{eq:patchwise_indep_posterior} and of the predictive PDFs~\eqref{eq:dynamical_mdl_a_iid_px},~\eqref{eq:dynamical_mdl_psi_iid_px}, this term can be written as:
\begin{align}
    & \Ex_{q_{\phi}(\bc_{t-1},\bpsi_{t-1}|\underline\by)} \big\{\KL\big(
    \nonumber\\ &
    q_{\phi}(\bc_{t},\bpsi_{t}|\bc_{t-1},\bpsi_{t-1},\underline\by) \big\| p_{\theta}(\bc_{t},\bpsi_{t}|\bc_{t-1},\bpsi_{t-1})\big) \big\}
    \nonumber\\ &
    = \sum_{n=1}^N \Ex_{q_{\phi}(\bc_{n,t-1},\bpsi_{n,t-1}|\underline\by_n)} \big\{ \KL\big(
    \\ &
    q_{\phi}(\bc_{n,t},\bpsi_{n,t}|\bc_{n,t-1},\bpsi_{n,t-1},\underline\by_{n}) \big\| p_{\theta}(\bc_{n,t},\bpsi_{n,t}|\bc_{t-1},\bpsi_{t-1})\big) \big\}.
    \nonumber
\end{align}
Combining these results, we can write the cost function $\calL(\theta,\phi,\underline\by)$ as in equation~\eqref{eq:pixelntime_factorized_ELBO} (depicted on top of the next page). Details on the computation of the log-likelihood and KL divergences can be found in Appendix~\ref{sec:appendix_KLdivs}.

\begin{table*}
\normalsize
\centering
\begin{align}
    \calL(\theta,\phi, & \underline\by) =
    \sum_{t=1}^T \sum_{n=1}^N \Ex_{q_{\phi}(\bc_{n,t},\bpsi_{n,t}|\underline\by_{n})} \big\{\log p_{\theta}(\by_{n,t}|\bc_{n,t},\bpsi_{n,t})\big\} 
    - \sum_{n=1}^N \KL\big(q_{\phi}(\bc_{n,0},\bpsi_{n,0}|\underline\by_{n}) \big\| p_{\theta}(\bc_{n,0},\bpsi_{n,0})\big)
    \nonumber\\
    & - \sum_{t=1}^T\sum_{n=1}^N \Ex_{q_{\phi}(\bc_{n,t-1},\bpsi_{n,t-1}|\underline\by_{n})} \Big\{\KL\big(q_{\phi}(\bc_{n,t},\bpsi_{n,t}|\bc_{n,t-1},\bpsi_{n,t-1},\underline\by_{n}) \big\| p_{\theta}(\bc_{n,t},\bpsi_{n,t}|\bc_{n,t-1},\bpsi_{n,t-1})\big) \Big\} \,.
    \label{eq:pixelntime_factorized_ELBO}
\end{align}
\vspace{-0.5cm}
\end{table*}

\begin{figure}[t]
    \centering
    \includegraphics[width=0.7\linewidth]{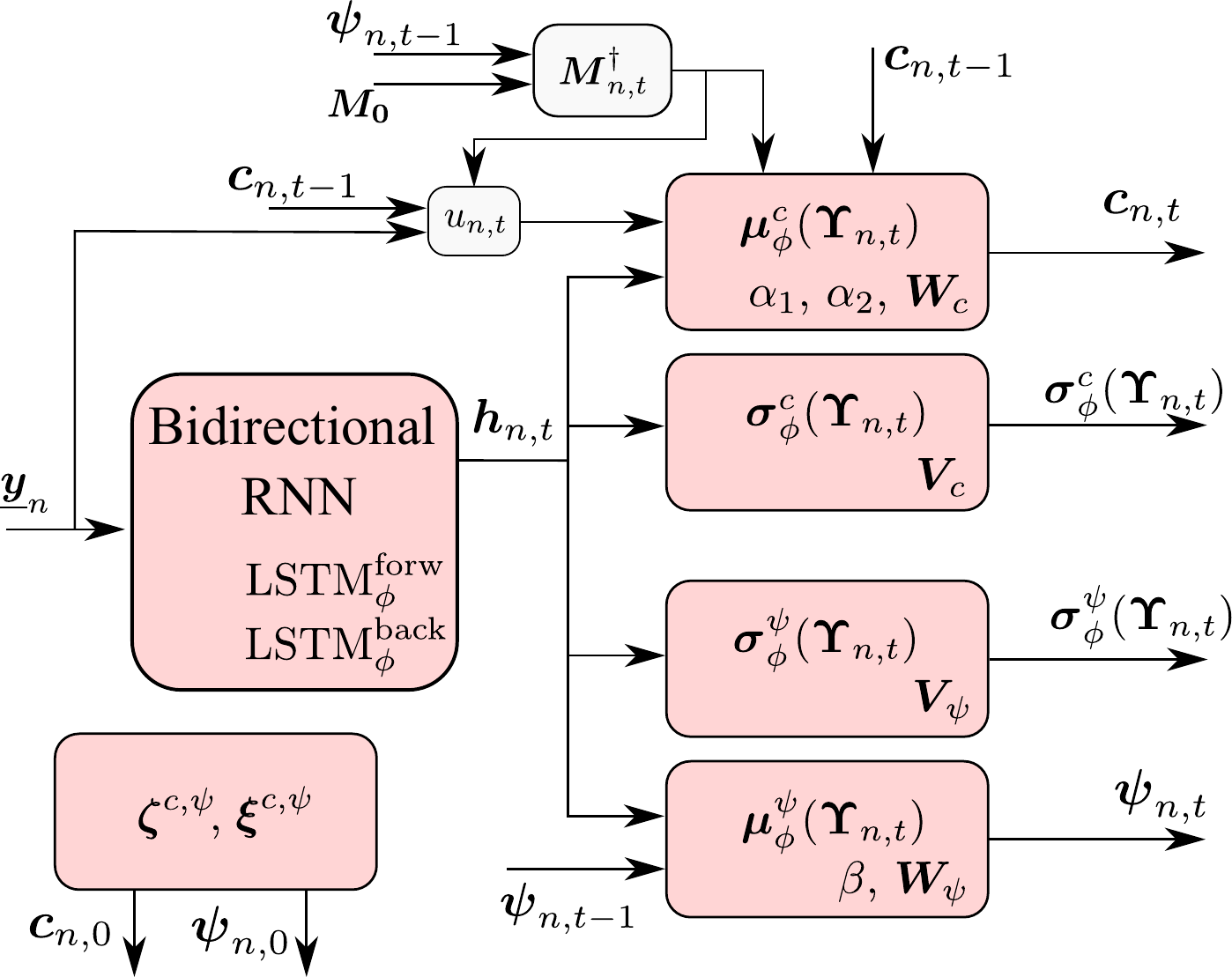}
    \vspace{-1.5ex}
    \caption{Diagram of the proposed network implementing the posteriors $q_{\phi}(\bc_{n,t},\bpsi_{n,t}|\bc_{n,t-1},\bpsi_{n,t-1},\underline\by_{n})$ and $q_{\phi}(\bc_0,\bpsi_0|\underline\by)$.}
    \label{fig:network_diagram}
\end{figure}

\subsection{An RNN-based implementation} 
\label{ssec:rnn_implementation}

A key question is how to define the functions $\bmu_{\phi}^{c}$, $\bmu_{\phi}^{\psi}$, $\bsigma_{\phi}^{c}$ and $\bsigma_{\phi}^{\psi}$ in~\eqref{eq:posterior_fact2} and~\eqref{eq:split_mean_cov_a_psi}, which parameterize the approximate posterior distribution. On the one hand, these have to be flexible to be able to approximate the true posterior, which cannot be written in the form of a simple and well-known distribution.
On the other hand, it is important to incorporate information from a physical modeling of the problem to make inference process more efficient, interpretable and stable.
Thus, we will parameterize the variational posterior distribution
using a lightweight RNN and, whenever possible, leveraging physically motivated models.

First, a bidirectional RNN is used to compute a set of feature-based representations, denoted by $\bh_{n,t}\in\amsmathbb{R}^H$, from the image pixel sequence $\underline\by_n$. In particular, we compute $\bh_{n,t}$ by combining the hidden states learned by two LSTMs~\cite{hochreiter1997lstm}:
\begin{align}
    \bh_{n,t}^{\forw} &= \operatorname{LSTM}_{\phi}^{\forw}(\bh_{n,t-1}^{\forw},\by_{n,t}), \,\,\, t=1,\ldots,T \,,
    \label{eq:RNN_h_forw}
    \\
    \bh_{n,t}^{\back} &= \operatorname{LSTM}_{\phi}^{\back}(\bh_{n,t+1}^{\back},\by_{n,t}), \,\,\, t=T,\ldots,1 \,,
    \label{eq:RNN_h_back}
    \\
    \bh_{n,t} &= \frac{1}{2}\big(\bh_{n,t}^{\forw} + \bh_{n,t}^{\back}\big) \,,
    \label{eq:RNN_h_avg}
\end{align}
for $n=1,\ldots,N$, where $\operatorname{LSTM}_{\phi}^{\forw}$ and $\operatorname{LSTM}_{\phi}^{\back}$ denote two LSTMs which process the data forward and backwards in time, respectively; their hidden state representation being given by $\bh_{n,t}^{\forw}$ and $\bh_{n,t}^{\back}$. We choose LSTMs due to their excellent performance in various sequence modeling tasks~\cite{lipton2015criticalReviewRNNs}. Moreover, a bidirectional RNN (i.e., two LSTMs) is used because at every time instant the posterior in~\eqref{eq:posterior_fact2} depends on the HI pixels at all time instants, $\underline\by_n$, whereas the LSTMs in~\eqref{eq:RNN_h_forw} and~\eqref{eq:RNN_h_back} depend only on past and future data, respectively.

The dimension of the RNN representation is selected as $H=(K+1)P$ (i.e., the dimension of the state vector). The gating units of the LSTMs use the sigmoid nonlinearity, while the input and hidden state units use the uses the hyperbolic tangent nonlinearity. Note that the parameters of these LSTMs will also be learned during inference by SDG using backpropagation through time~\cite{lipton2015criticalReviewRNNs}.

We now use the representation $\bh_{n,t}$ to parameterize $\bmu_{\phi}^{c}$, $\bmu_{\phi}^{\psi}$, $\bsigma_{\phi}^{c}$ and $\bsigma_{\phi}^{\psi}$. To introduce physical knowledge, we follow the general idea of using hybrid models~\cite{li2021modelBasedAECsSU,imbiriba2022hybrid}, in which an approximate model is complemented by a learnable component (in this case derived from the RNN).
In particular, for the posterior mean of the abundances, $\bmu_{\phi}^{c}$, we construct an approximate model by assuming that 1) a least squares solution provides a crude abundance estimate, 2) the abundances are temporally smooth but may undergo sudden changes, and 3) abrupt abundance changes lead to abrupt changes in the pixels.
For the variability parameters, $\bmu_{\phi}^{\psi}$, we consider it to be temporally smooth.
For the standard deviations, $\bsigma_{\phi}^{c}$ and $\bsigma_{\phi}^{\psi}$, we don't have a good physical model; thus, we use a purely non-parametric representation. 
In the following, we define each of these functions explicitly; an illustrative diagram can be seen in Figure~\ref{fig:network_diagram}.

Considering the RNN features $\bh_{n,t}$, the abundance posterior means $\bmu_{\phi}^{c}$ is then parameterized as:
\begin{align}
    \bmu_{\phi}^{c}(\bUpsilon_{n,t}) ={} & \bpi^{-1}\Big(\alpha_1 (1-u_{n,t}) \bpi(\bc_{n,t-1}) 
    \nonumber \\
    & + \alpha_2 u_{n,t}\big(\bM_{n,t-1}^\dagger\by_{n,t} + \bW_{\!c} \bh_{n,t}\big) \Big) \,,
    \label{eq:parametrization_mu_c}
\end{align}
where $\bM_{n,t-1}=\bM_0\odot(\mathbb{1}+\bD\vect^{-1}(\bpsi_{n,t-1}))$ is the predicted EM matrix at pixel $n$ and time $t-1$, $\alpha_1$ and $\alpha_2$ are trainable, real-valued weighting coefficients, and $\bW_{\!c}\in\amsmathbb{R}^{P\times (K+1)P}$ is a trainable matrix that maps the hidden RNN representations to the abundances in the softmax basis. 
The scalar coefficient $u_{n,t}\in[0,1]$, defined as $u_{n,t}=\frac{1}{2P}\|s(\bM_{n,t-1}^\dagger\by_{n,t})-\bpi(\bc_{n,t-1})\|_1$, measures the difference between the predicted abundances $\bpi(\bc_{n,t-1})$ and a crude estimation of the current abundances at time $t$, given by $s(\bM_{n,t-1}^\dagger\by_{n,t})$, where the fixed function $s(\cdot)$ projects the linear regression solution $\bM_{n,t-1}^\dagger\by_{n,t}$ to the unit simplex.
Thus, $u_{n,t}$ works as a crude abrupt change detector.

The parametrization~\eqref{eq:parametrization_mu_c} can be seen as a weighted combination of three terms: the abundances at the previous time instant, a crude abundance estimate at time $t$ computed by linear regression, and a non-parametric term depending on $\bh_{n,t}$. The balance between them depends on the trainable weights and on the change detector $u_{n,t}$.
When there are no changes, $u_{n,t}$ is small, which gives a higher contribution to the predicted abundances $\bc_{n,t-1}$ in~\eqref{eq:parametrization_mu_c}. On the other hand, if there is an abrupt change, $u_{n,t}$ is large, giving a higher contribution to the sum of the last two terms in~\eqref{eq:parametrization_mu_c}, which is a linear regression-based abundance estimate augmented by a non-parametric RNN-based representation. This parametrization is particularly relevant since the generative model~\eqref{eq:abundances_time_evol_softmax} does not explicitly represent abrupt changes.

For the function $\bmu_{\phi}^{\psi}$, to leverage temporal smoothness we consider a weighted linear combination of the variability coefficients at the previous time instant $\bpsi_{n,t-1}$ and a linear mapping of the hidden RNN representation:
\begin{align}
    \bmu_{\phi}^{\psi}(\bUpsilon_{n,t}) ={} & \beta \bpsi_{n,t-1} + \bW_{\!\psi}\bh_{n,t} \,,
    \label{eq:parametrization_mu_psi}
\end{align}
where $\beta$ is a real-valued weight, and $\bW_{\!\psi}\in\amsmathbb{R}^{KP\times (K+1)P}$ is a matrix that computes the variability coefficients' innovation from the RNN representation $\bh_{n,t}$, both of which are trainable.
Note that by not considering abrupt changes to occur in $\bpsi_{t}$ we obtain a simpler model compared to $\bmu_{\phi}^{c}$.

The standard deviations $\bsigma_{\phi}^{c}$ and $\bsigma_{\phi}^{\psi}$ are computed based on a fully non-parametric model, which is given as linear mappings of the RNN representations $\bh_{n,t}$:
\begin{align}
    \bsigma_{\phi}^{c}(\bUpsilon_{n,t}) & = \exp\big(\bV_{\!c} \bh_{n,t}\big) \,,
    \\
    \bsigma_{\phi}^{\psi}(\bUpsilon_{n,t}) & = \exp\big(\bV_{\!\psi} \bh_{n,t}\big) \,,
\end{align}
where $\bV_{\!c}\in\amsmathbb{R}^{P\times (K+1)P}$ and $\bV_{\!\psi}\in\amsmathbb{R}^{KP\times (K+1)P}$ are the transformation matrices, and the exponential function is applied elementwise in order to ensure nonnegativity of the standard deviations. The parameters of the approximate posterior are finally denoted by $\phi=\big\{\bzeta^{c,\psi},$
$\bxi^{c,\psi},\operatorname{LSTM}_{\phi}^{\forw},\operatorname{LSTM}_{\phi}^{\back},\alpha_1,\alpha_2,\beta,\bW_{\!c},\bV_{\!c},\bW_{\!\psi},\bV_{\!\psi}\big\}$. Note that all parameters in $\phi$ will be learned using SGD.

\paragraph*{\textbf{Approximating the expectations and optimization}}

To optimize~\eqref{eq:pixelntime_factorized_ELBO} using stochastic backpropagation, it is necessary to estimate gradients of expectations whose distribution depend on $\theta$ and $\phi$, which are the parameters to be optimized. To address this issue, we consider the reparametrization trick, which provides low-variance gradient estimates~\cite{kingma2014VAEs}. %
This is performed by writing the random variables inside the expectations as deterministic functions of a random variable that does not depend on $\phi$. In general, for a distribution $q_{\phi}(\bx)$ and function $f$, this can be formulated as
$\Ex_{q_{\phi}(\bx)} \{f(\bx)\} = \Ex_{p(\bepsilon)} \{f(g(\bepsilon))\}$,
where $g$ is a function such that $\bx$ and $g(\bepsilon)$ have the same distribution, and $p(\bepsilon)$ does not depend on $\phi$.
Applying this to the expectations in~\eqref{eq:pixelntime_factorized_ELBO} and considering that the posterior in our model is Gaussian, this is achieved as:
\begin{align}
    \big[\ba_{n,t}^\top,\bpsi_{n,t}^\top\big]^\top &= \bmu_{\phi}^{c,\psi}(\bUpsilon_{n,t}) + \bsigma_{\phi}^{c,\psi}(\bUpsilon_{n,t}) \odot \bepsilon \,,
    \label{eq:reparametrization_1}
\end{align}
for $t=1,\ldots,T$, where $\bepsilon\sim\calN(\cb{0},\bI)$. It can be verified that the random variables in~\eqref{eq:reparametrization_1} are sampled according to the distribution $q_{\phi}(\bc_{n,t},\bpsi_{n,t}|\underline\by_{n})$. 
Thus, by using this reparametrization, the expectations in~\eqref{eq:pixelntime_factorized_ELBO} can be rewritten in terms of expectations of $\bepsilon$, which we subsequently approximate using a one-sample Monte Carlo estimate and denote by $\widehat{\calL}(\theta,\phi,\underline\by)$.

The approximated cost function $\widehat{\calL}(\theta,\phi,\underline\by)$ is then optimized with respect to both $\theta$ and $\phi$ (i.e., the parameters of the generative model and of the variational posterior) using the Adam stochastic optimization method~\cite{kingma2014adam}.
We used a learning rate of $0.001$ and a batch size of $128$. Training was performed for $30$ epochs.
The full MTSU process performed by ReDSUNN is summarized in Algorithm~\ref{alg:alg1}.

Since the cost function is non-convex, the initialization of the parameters can have an important impact on the solution. The parameters of the neural networks $\sigma_a$, $\operatorname{LSTM}_{\phi}^{\forw}$, $\operatorname{LSTM}_{\phi}^{\back}$, and the matrices $\bV_{\!c}$ and $\bV_{\!\psi}$ are initialized randomly using Glorot initialization~\cite{glorot2010understandingInitializeNNs}. $\bW_{\!c}$ and $\bW_{\!\psi}$ are initialized with zeros, and $\beta=\alpha_1=\alpha_2=1$.
$\bM_0$ was \cblue{initialized using the vertex component analysis (VCA) algorithm~\cite{Nascimento2005}}, and
$\sigma_r=0.0001$ (corresponding to an SNR of about 35dB for spectra with standard deviation $0.5$).
For the parameters of the initial prior and variational posterior PDFs, we initialized the means $\bnu_0^a$, $\bnu_0^\psi$ and $\bzeta^{c,\psi}$ with zeros, and the variances $\bgamma_0^a$, $\bgamma_0^\psi$ and $\bxi^{c,\psi}$ with ones, making the initial PDFs standard Gaussians.

\begin{algorithm} [!t]
\footnotesize
\SetKwInOut{Input}{Input}
\SetKwInOut{Output}{Output}
\caption{ReDSUNN}\label{alg:alg1}
\Input{HIs $\by_{1},\ldots,\by_{T}$, hyperparameters $P$, $K$ and $\sigma_{\psi}$.}

Initialize $\theta_0$ and $\phi_0$ as described in Section~\ref{ssec:rnn_implementation} \;

Use Adam~\cite{kingma2014adam} to maximize $\widehat{\calL}(\theta,\phi,\underline\by)$ w.r.t. both $\phi$ and $\theta$ \;

\For{$n=1,\ldots,N$}{
Compute $\widehat{\bc}_{n,0},\widehat{\bpsi}_{n,0}$ as the means of $q_{\phi}(\bc_{n,0},\bpsi_{n,0}|\underline\by_{n})$ using~\eqref{eq:posterior_fact2_init}\;

Compute the $\bh_{n,1},\ldots,\bh_{n,T}$ using~\eqref{eq:RNN_h_forw},~\eqref{eq:RNN_h_back} and~\eqref{eq:RNN_h_avg} \;

\For{$t=1,\ldots,T$}{
Compute $\widehat{\bc}_{n,t},\widehat{\bpsi}_{n,t}$ as the means of $q_{\phi}(\bc_t,\bpsi_t|\widehat{\bc}_{t-1},\widehat{\bpsi}_{t-1},\underline\by)$ using~\eqref{eq:parametrization_mu_c} and~\eqref{eq:parametrization_mu_psi} \;
}
}
Set $\widehat{\ba}_{n,t}=\bpi(\widehat{\bc}_{n,t})$, $\widehat{\bM}_{n,t}=\widehat{\bM}_0\odot(\mathbb{1}+\bD\vect^{-1}(\widehat{\bpsi}_{n,t}))$ \;

\Output{$\widehat{\ba}_{n,t}$, $\widehat{\bM}_{n,t}$, for $n=1,\ldots,N$ and $t=1,\ldots,T$.} 
\end{algorithm}

\subsection{Model complexity and comparisons}
\label{sec:complexityComparisons}

We now summarize the parameters of the generative model, of the variational posterior, and their dimensionality (i.e., the number of parameters that have to be inferred). This can be seen in Table~\ref{tab:number_params}. To compute the number of parameters corresponding to the LSTMs, we note that each LSTM has four input-hidden weight matrices, four hidden-hidden weight matrices, and four biases (where the input is of size $L$, and the hidden state of size $(K+1)P$). It is instructive to compare the amount of parameters to other methods in the literature. By using a Markovity assumption, a shared posterior distribution for all pixels, and an RNN posterior parametrization, the amount of parameters to be learned by ReDSUNN in Table~\ref{tab:number_params} does not scale with either $N$ or $T$, differently from previous methods such as OU~\cite{thouvenin2016online} or the HBUN~\cite{liu2021bayesianSU_multitemporal}.

\begin{table}[h]
\footnotesize
\caption{Variables to be estimated and number of parameters.}
\vspace{-0.3cm}
\centering
\renewcommand{\arraystretch}{1.2}
\begin{tabular}{l|ccccccccccc}	
\bottomrule
\multicolumn{2}{c}{Generative model ($\theta$)}	\\
\toprule
$\bM_0$ & $LP$ \\
$\sigma_r$ & 1 \\
$\sigma_a$ & $P(P+1)R_{\sigma_a}$ \\
$\bnu_0^a,\bnu_0^\psi,\bgamma_0^a,\bgamma_0^\psi$ & $2(K+1)P$ \\

\bottomrule
\multicolumn{2}{c}{Variational posterior ($\phi$)}	\\
\toprule
$\operatorname{LSTM}_{\phi}^{\forw}$, $\operatorname{LSTM}_{\phi}^{\back}$ & $8(K+1)P\big((K+1)P + L + 1\big)$ \\
$\alpha_1$, $\alpha_2$, $\beta$ & $3$ \\
$\bW_{\!c}$, $\bV_{\!c}$ & $2(P^2(K+1))$ \\
$\bW_{\!\psi}$, $\bV_{\!\psi}$ & $2(KP^2(K+1))$\\
$\bzeta^{c,\psi}$, $\bxi^{c,\psi}$ & $2(K+1)P$ \\
\midrule
\end{tabular}
\label{tab:number_params}
\end{table}

\section{Results}
\label{sec:results}

The performance of the proposed ReDSUNN algorithm is evaluated using simulations with synthetic and real data. We compare our method with the fully constrained least squares (FCLS), online unmixing (OU)~\cite{thouvenin2016online}, HBUN~\cite{liu2021bayesianSU_multitemporal}, and with a Kalman filter and expectation maximization-based strategy (referred to simply as \emph{Kalman})~\cite{borsoi2020multitemporalUKalmanEM}.
The EMs used by FCLS were extracted by the VCA algorithm at each time instant~\cite{Nascimento2005}. The reference EMs required by the Kalman method, and the initialization for the EMs in OU, HBUN and for the proposed method were all set with the signatures obtained by applying VCA to the matrix $[\by_{1,1},\ldots,\by_{n,t}\ldots,\by_{N,T}]\in\amsmathbb{R}^{L\times NT}$ formed by concatenating the HI pixels for all time instants.

The abundances and EM scaling factors estimated by \mbox{ReDSUNN} are set according to Algorithm~\ref{alg:alg1}.
The hyperparameters of all algorithms were adjusted so as to obtain high abundance reconstruction performance. For ReDSUNN, parameters $K$ and $\sigma_{\psi}$ (which are not optimized) were searched within the ranges $K\in\{1,\ldots,10\}$ and $\sigma_{\psi}\in\{10^{-5},\ldots,0.1,1\}$, and $R_{\sigma_a}=2$ layers were used to parameterize function $\sigma_a(\cdot)$ in~\eqref{eq:abundances_time_evol_softmax}.
For the other algorithms, their parameters were selected in the ranges indicated in their original publications.
The proposed method was implemented in Pytorch (codes will be available at \url{https://github.com/ricardoborsoi/ReDSUNN}). The remaining methods were implemented in Matlab (codes were provided by the original authors). All experiments were run in a desktop computer with an Intel Xeon\textsuperscript{TM} W-2104 CPU with four 3.2GHz cores and 24GB of RAM. No GPU was used in the simulations. ReDSUNN, OU and Kalman used parallelization in their implementations.

The quantitative performance of the algorithms was evaluated using the average normalized mean squared error (NRMSE), between the abundances, EMs, and reconstructed HIs, which are computed as 
$\text{NRMSE}_{\bA} = \big(\frac{1}{T} \sum_{t=1}^T\sum_{n=1}^N \|\ba_{n,t}-\widehat{\ba}_{n,t}\|^2\big/\|\ba_t\|^2\big)^{1/2}$, $\text{NRMSE}_{\bM} = \big(\frac{1}{NT} \sum_{t=1}^T \sum_{n=1}^N \|\bM_{n,t}-\widehat{\bM}_{n,t}\|_F^2\big/\|\bM_{n,t}\|_F^2\big)^{1/2}$, and
$\text{NRMSE}_{\bY} = \big(\frac{1}{T} \sum_{t=1}^T\sum_{n=1}^N \big\|\by_{n,t}-\widehat{\bM}_{n,t}\widehat{\ba}_t\big\|^2\big/\|\by_t\|^2\big)^{1/2}$,
where $\widehat{\ba}_{n,t}$ and $\widehat{\bM}_{n,t}$ denote the estimated abundances and EMs.
To evaluate the EMs, we also computed the average spectral angle mapper (SAM) as
$\text{SAM}_{\bM} = \frac{1}{TNP} \sum_{t=1}^{T} \sum_{n=1}^{N} \sum_{j=1}^P \arccos \Big( \frac{\bm_{n,t,j}^\top\widehat{\bm}_{n,t,j}}{\|\bm_{n,t,j}\|\|\widehat{\bm}_{n,t,j}\|} \Big)$,
in which $\bm_{n,t,j}$ and $\widehat{\bm}_{n,t,j}$ are the true and estimated EM signatures for time $t$, pixel $n$ and EM $j$.

\begin{table}[t] %
\footnotesize
\caption{Quantitative results of the simulations using synthetic data.}
\vspace{-2.5ex}
\centering
\renewcommand{\arraystretch}{1.2}
\setlength{\tabcolsep}{3.3pt}
\begin{tabular}{lccccr}
\bottomrule
& $\text{NRMSE}_{\bA}$ & $\text{NRMSE}_{\bM}$ & $\text{SAM}_{\bM}$ & $\text{NRMSE}_{\bY}$ & Time \\
\toprule	
\bottomrule
\multicolumn{6}{c}{Data Sequence 1 -- DS1} \\
\toprule%
FCLS	&	0.537	&	--	&	--	&	0.086	&	\textbf{2.7}	\\
OU  	&	0.434	&	0.342	&	0.260	&	0.051	&	24.9	\\
HBUN	&	0.479	&	0.355	&	0.162	&	\textbf{0.050}	&	542.6	\\
Kalman	&	0.356	&	0.124	&	0.076	&	0.061	&	2422.8	\\
ReDSUNN	&	\textbf{0.318}	&	\textbf{0.117}	&	\textbf{0.075}	&	0.089	&	479.0	\\

\bottomrule
\multicolumn{6}{c}{Data Sequence 2 -- DS2}	\\
\toprule%
FCLS	&	0.500	&	--	&	--	&	0.122	&	\textbf{7.3}	\\
OU	    &	0.335	&	0.256	&	\textbf{0.120}	&	0.055	&	60.6	\\
HBUN	&	0.474	&	0.515	&	0.141	&	\textbf{0.050}	&	2166.0	\\
Kalman	&	0.659	&	12.222	&	0.496	&	0.108	&	5937.4	\\
ReDSUNN	&	\textbf{0.294}	&	\textbf{0.203}	&	0.289	&	0.160	&	1231.3	\\

\toprule		
\end{tabular}
\label{tab:results_synthData}
\end{table}

\begin{figure}
    \centering
    \includegraphics[width=0.9\linewidth]{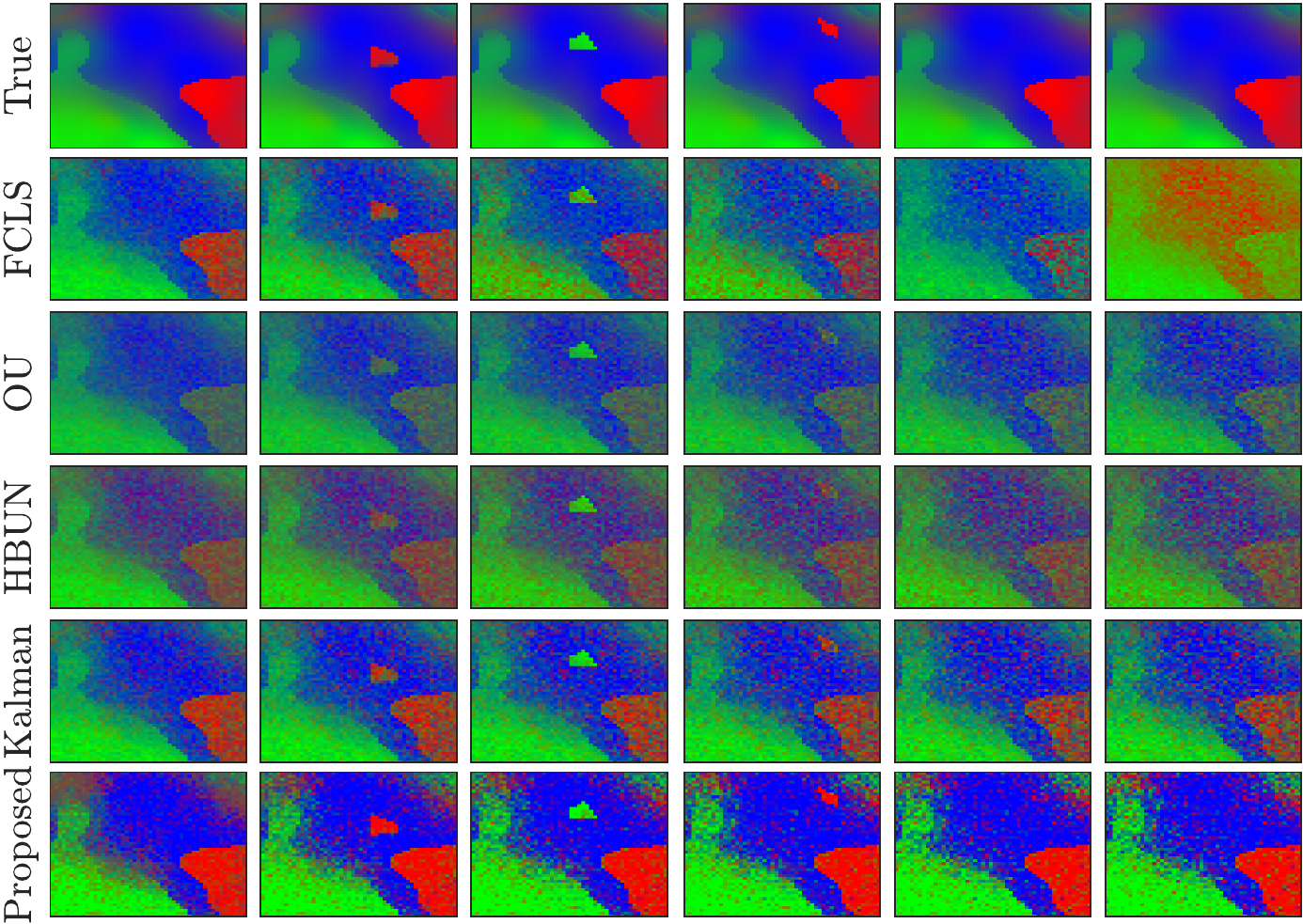}
    \vspace{-1.5ex}
    \caption{Estimated abundance maps and ground truth for DS1, shown as composite color maps (that is, the abundances of the EMs \#1, \#2 and \#3 are represented as the red, green, and blue color channels).}
    \label{fig:synth_ex1_abunds}
\end{figure}

\subsection{Simulations with synthetic data}

\begin{figure}[!t]
    \centering
    {\tiny \hspace{0.02\linewidth} EM \#1 \hspace{0.2\linewidth} EM \#2 \hspace{0.2\linewidth} EM \#3} \\
    \vspace{0.2ex}
    \includegraphics[width=0.8\linewidth]{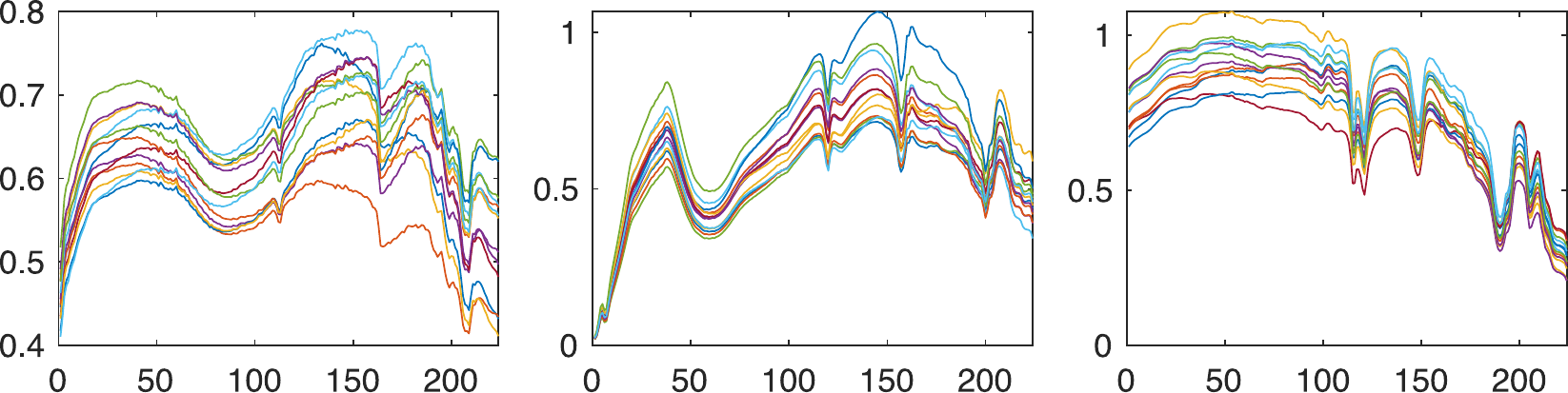}\\ \smallskip
    \includegraphics[width=0.8\linewidth]{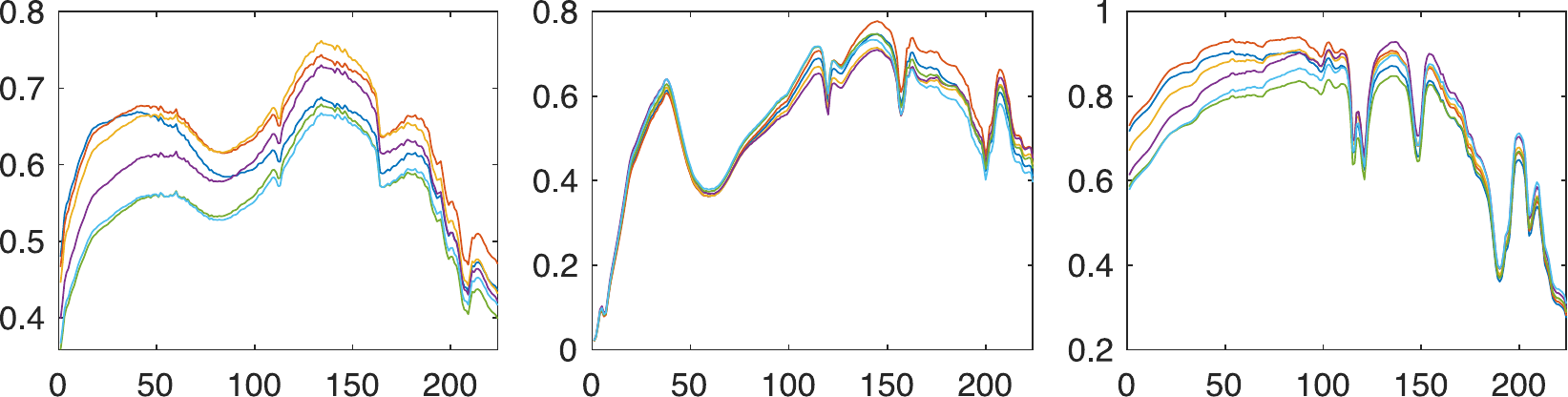}\\
    \vspace{-1.5ex}
    \caption{True EMs for the DS1, sampled over space, for time instant $t=3$ (top), and over time, for pixel $n=1$ (bottom).}
    \label{fig:endmembers_ex1_true}
    {\tiny \hspace{0.02\linewidth} EM \#1 \hspace{0.2\linewidth} EM \#2 \hspace{0.2\linewidth} EM \#3} \\
    \vspace{0.4ex}
    \includegraphics[width=0.8\linewidth]{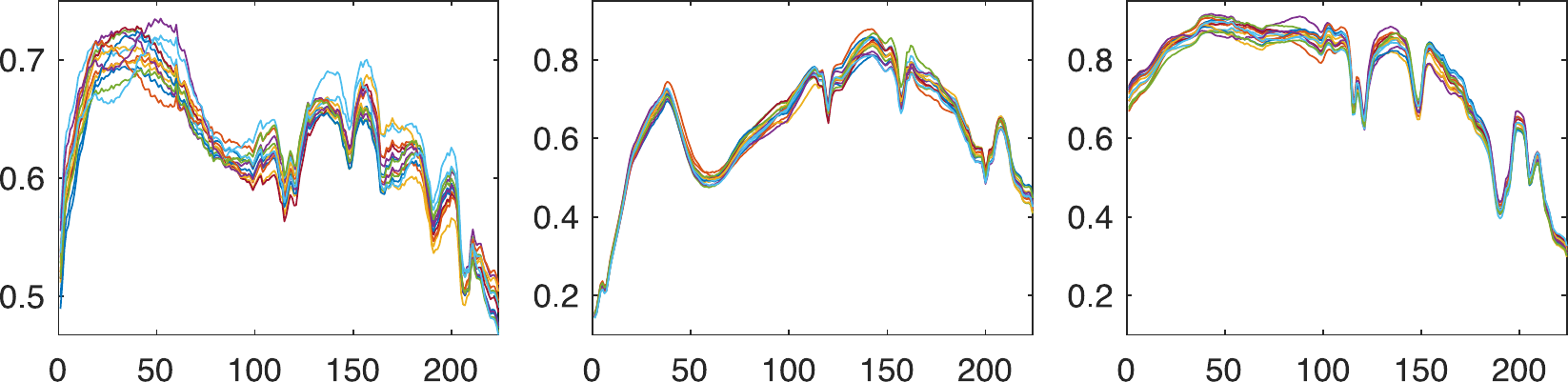}\\ \smallskip
    \includegraphics[width=0.8\linewidth]{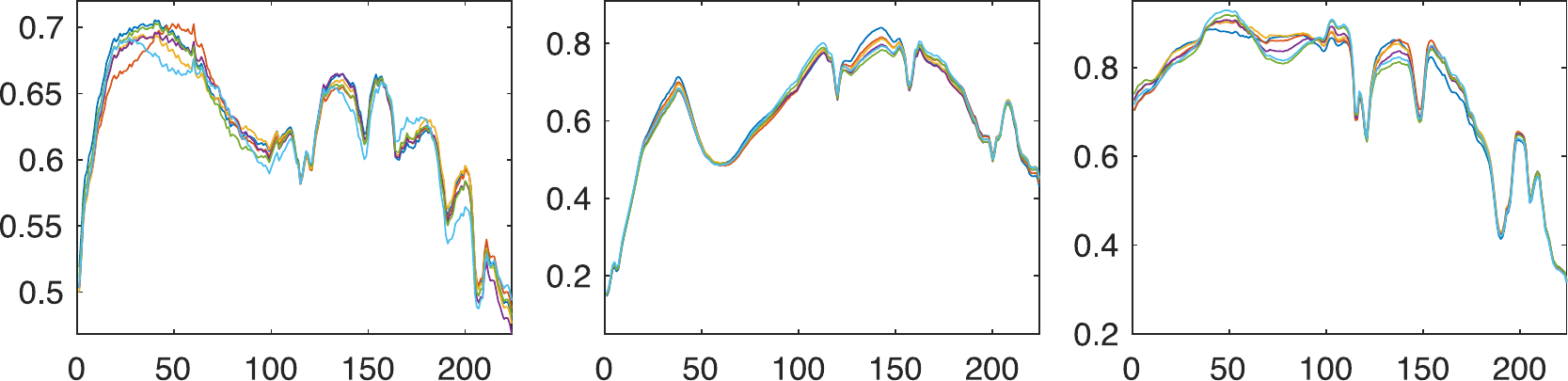}\\
    \vspace{-1.5ex}
    \caption{Estimated EMs for the DS1, sampled over space, for time instant $t=3$ (top), and over time, for pixel $n=1$ (bottom).}
    \label{fig:endmembers_ex1_est}
\end{figure}

Two synthetic datasets were considered with spatiotemporal abundance and EM variability. The first dataset, referred to as Data Sequence 1 (DS1), contains $P=3$ EMs and $T=6$ HIs. The HIs are generated from sequences of abundance maps with $N=50\times50$ pixels containing localized abrupt changes for $t\in\{2,3,4,5\}$ (depicted in the first row of Figure~\ref{fig:synth_ex1_abunds}). The EMs for each pixel and time instant were generated as follows. First, three signatures with $L=224$ bands were selected from the USGS library and used as a reference EM matrix. Then, for the first time instant ($t=1$), spatial EM variability was introduced by following the model in~\cite{Thouvenin_IEEE_TSP_2016_PLMM}, in which the EMs in each pixel ($\bM_{n,1}$, $n=1,\ldots,N$) were generated by multiplying the reference signatures with piecewise linear random scaling factors with amplitude in the interval $[0.85,1.15]$.
For each subsequent time instant $t>1$, the EMs were also generated as scaled versions of the reference spectral signatures. However, to introduce temporal EM variability, the scaling factors at time $t$ are defined to be the sum of the scaling factors at time $t-1$ plus random piecewise linear functions in the range $[-0.1,0.1]$. Samples of the generated EMs can be seen in Figure~\ref{fig:endmembers_ex1_true}. These EM matrices $\bM_{n,t}$ are then used to generate the HI pixels using the LMM \eqref{eq:LMM}, with the measurement noise $\br_{n,t}$ being white and Gaussian with an SNR of 30~dB.
The second dataset, referred to as Data Sequence 2 (DS2), contained $P=4$ EMs and $N=50\times50$ pixels. A sequence of abundance maps generated randomly according to a Gaussian random field and containing small, spatially compact abrupt changes was considered to generate $T=15$ HIs.
To introduce realistic spectral variability, the EM signatures at each pixel and time instant were randomly selected from a set of pure pixels of water, vegetation, soil and road that were manually extracted from the Jasper Ridge HI, with $L=198$ bands. The HI sequence was then generated according to the multitemporal LMM~\eqref{eq:LMM}, with the $\br_{n,t}$ being white Gaussian noise with an SNR of 30~dB. The parameters of the ReDSUNN were $K=10$ and $\sigma_{\psi}=10^{-5}$ for DS1, and $K=2$ and $\sigma_{\psi}=10^{-5}$ for DS2.
The quantitative results are presented in Table~\ref{tab:results_synthData}, while the visual results (only shown for DS1 due to space limitations) are depicted in Figures~\ref{fig:synth_ex1_abunds} and~\ref{fig:endmembers_ex1_est}.

\subsubsection{Discussion}

It can be seen from Table~\ref{tab:results_synthData} that ReDSUNN achieved the best abundance estimation performance for both datasets. OU and HBUN achieved consistent but intermediate results, while the performance of the Kalman filter was good for DS1 but very poor for DS2. The FLCS, which does not take temporal information of spatial EM variability into account, did not perform very well, having the worse abundance reconstructions on average for both datasets. From the estimated abundances in Figure~\ref{fig:synth_ex1_abunds}, it can be seen that ReDSUNN's results are the closest to the ground truth. However, the results for all methods were relatively noisy. The abundances recovered by the Kalman filter, OU and HBUN indicated more heavily mixed pixels. FCLS achieves reasonable performance for $t\leq4$, but led to a completely wrong estimation for $t=6$. The changes occurring in the ground truth abundances can be observed in the estimations of all methods, although they are more clearly visible in the Kalman and ReDSUNN results since these methods led to a larger separation between the different materials.
The visual abundance results for DS2 (not shown due to space limitations) were qualitatively similar to those of DS1, with the exception that the Kalman filter failed to identify the soil EM for all images in the sequence, which explains its poor performance.

The ReDSUNN method also obtained the best EM estimation performance all metrics except for the SAM in DS2, in which OU achieved the best result followed by HBUN. The Kalman filter obtained good results for DS1 (close to ReDSUNN), but poor results in DS2. This happened despite the Kalman filter obtaining reasonable image reconstruction errors $\text{NRMSE}_{\bY}$ for both DS1 and DS2. Samples of the true and estimated EMs in Figures~\ref{fig:endmembers_ex1_true} and~\ref{fig:endmembers_ex1_est} (only shown for DS1 and for ReDSUNN due to space limitations) indicate that the EMs are correctly recovered. However, there are some differences, particularly in the shape of the first EM (which show higher amplitude in the ground truth compared to the estimates). Moreover, the amount of variability was lower in the retrieved EMs compared to the ground truth; 
this occurs for the synthetic examples since the hyperparameter $\sigma_{\psi}$, which controls the flexibility of the EM model, was selected to provide the best performance in terms of $\text{NRMSE}_{\bA}$, leading to relatively small $\sigma_{\psi}$ values. This is further illustrated in the experiments shown in~Section~\ref{ssec:ablation_exps}.

The lowest image reconstruction errors ($\text{NRMSE}_{\bY}$) were obtained by OU and HBUN, while those obtained by \mbox{ReDSUNN} were similar to those by FCLS. This is expected, since $\text{NRMSE}_{\bY}$ is closely related to the number of learnable parameters of each algorithm, and is not directly related to the abundance or EM reconstruction performance. This explains the higher reconstruction error by ReDSUNN since, as discussed in Section~\ref{sec:complexityComparisons}, the shared parametrization of the variational posterior PDF leads to a relatively low number of parameters, which also helps to mitigate overfitting.
\cblue{Nonetheless, for the synthetic data sequences (DS1 and DS2) ReDSUNN still has between 30\% and 50\% more learnable parameters than OU. Its parametrizaition becomes significantly more favorable when the images have a larger amounts of pixels, such as in the experiments with the Lake Tahoe images presented in Section~\ref{sec:exprtiments_tahoe}.}
The computation times show a clear separation between FCLS and OU, which were faster, and HBUN, the Kalman filter and ReDSUNN, which took longer to run. This indicates that the proposed method has a competitive computational performance when compared to more complex algorithms.

\begin{figure}[t]
    \centering
    \includegraphics[width=0.49\linewidth]{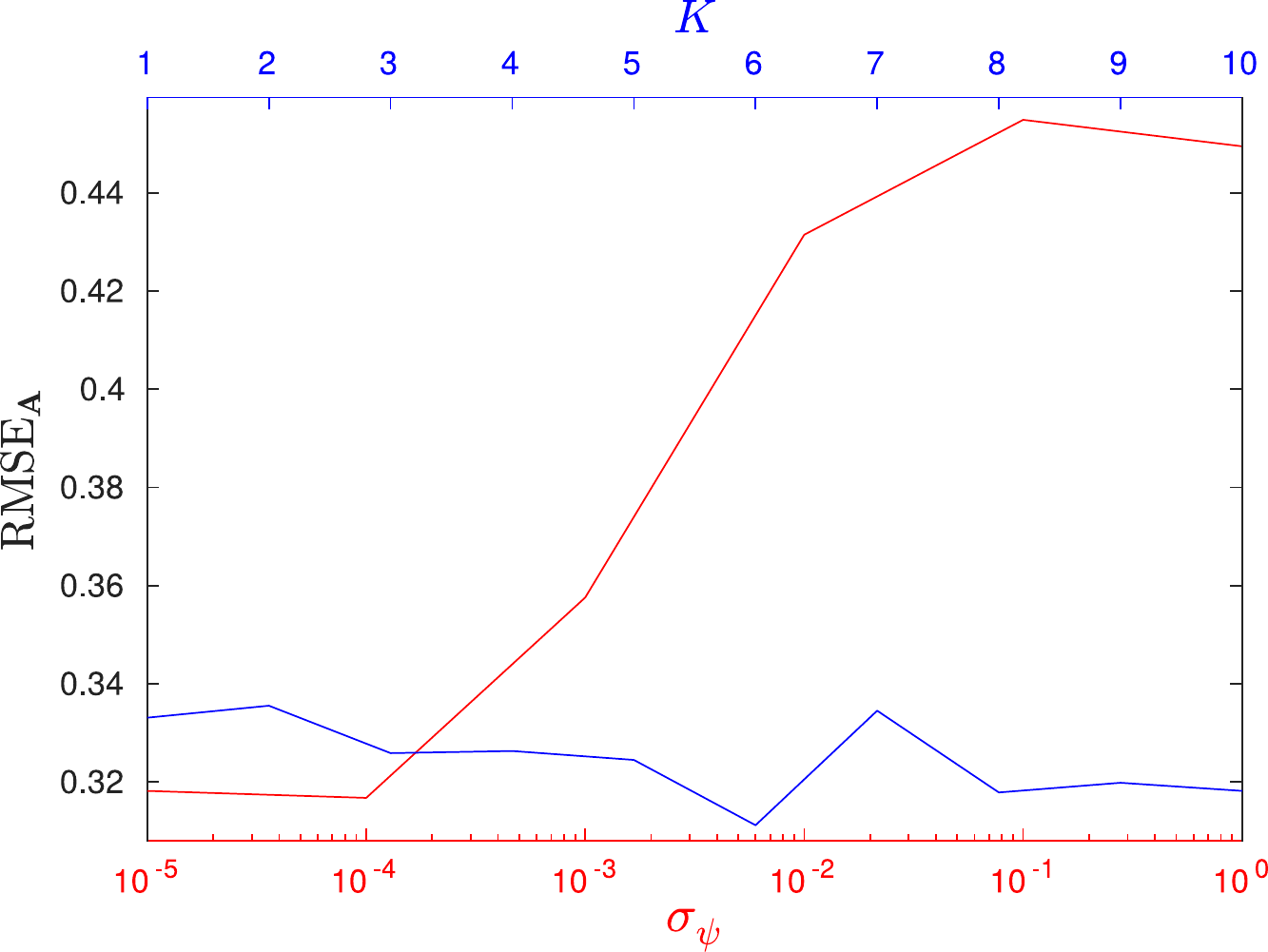}
    \includegraphics[width=0.49\linewidth]{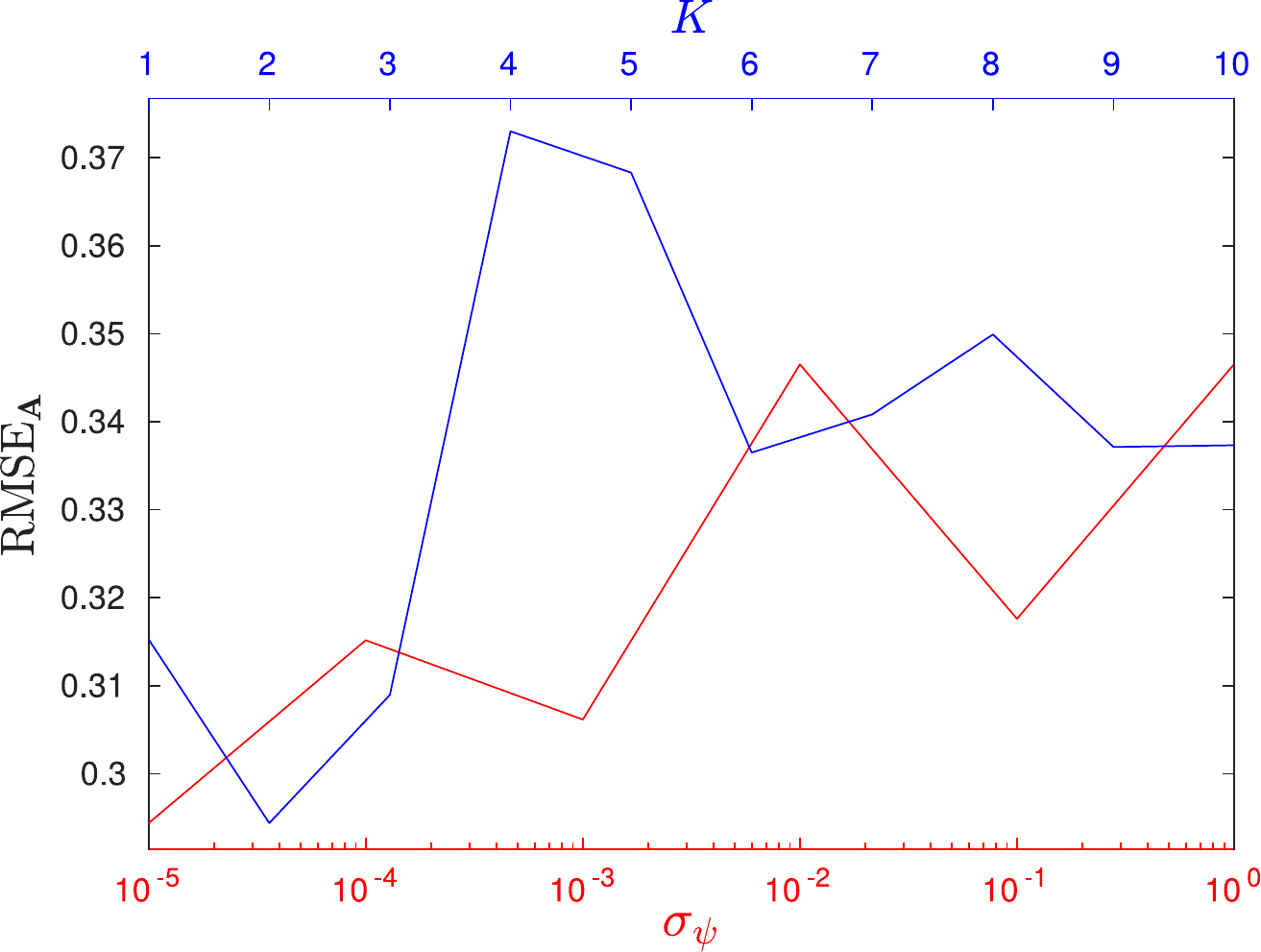}\\
    \vspace{-1.2ex}
    \caption{Abundance RMSE as a function of hyperparameters $K$ and $\sigma_{\psi}$ for DS1 (left) and DS2 (right).}
    \label{fig:ablation}
\end{figure}

\subsection{Sensitivity analysis}
\label{ssec:ablation_exps}

To measure the influence of different hyperparameters on the performance of the method, we evaluated how $\text{NRMSE}_{\bA}$ varied as a function of the hyperparameters, namely, the number of basis vectors for the variability model, $K$, and the innovation standard deviation of the EM variability parameters, $\sigma_{\psi}$. The results for both DS1 and DS2 can be seen in Figure~\ref{fig:ablation}.
It can be seen that for DS1, the performance of ReDSUNN is not heavily affected by the number of bases $K$ within the evaluated range. However, $\sigma_{\psi}$ has a larger impact on the result, with smaller values leading to a lower $\text{NRMSE}_{\bA}$. For DS2, smaller values for both parameters generally lead to lower $\text{NRMSE}_{\bA}$ results, although there performance varied more with $K$ and $\sigma_{\psi}$. For both datasets, small variations of these parameters around the optimal values lead to similar results.
In general, the larger the value of $\sigma_{\psi}$, the more temporal EM variability is allowed by the model, whereas the larger the value of $K$, the more complex the spatial and temporal EM variability the model can represent. Devising a methodology to automatically tune these parameters is an interesting question for future work.

\begin{table}[h]
\footnotesize
\caption{Quantitative results for the Lake Tahoe HI sequence.}
\vspace{-2.5ex}
\centering
\renewcommand{\arraystretch}{1.2}
\begin{tabular}{lccrccr}
\bottomrule
& FCLS & OU & HBUN & Kalman & ReDSUNN  \\
\toprule
$\text{NRMSE}_{\bY}$ & 
0.321 & 
0.058 &
\textbf{0.054} &
0.185 &
0.114 \\

Time & 
\textbf{16.1} &
92.5 &
3381.9 &
4607.7 &
2857.2 \\

\toprule		
\end{tabular}
\label{tab:results_realData}
\end{table}

\begin{figure}[t]
    \centering
    {\scriptsize 
    10--04--2014 \hspace{0.8ex}
    02--06--2014 \hspace{0.8ex}
    19--09--2014 \hspace{0.8ex}
    17--11--2014 \hspace{0.8ex}
    29--04--2015 \hspace{0.8ex}
    13--10--2015} \\
    \vspace{0.4ex}
    \includegraphics[width=\linewidth, height=0.17\linewidth]{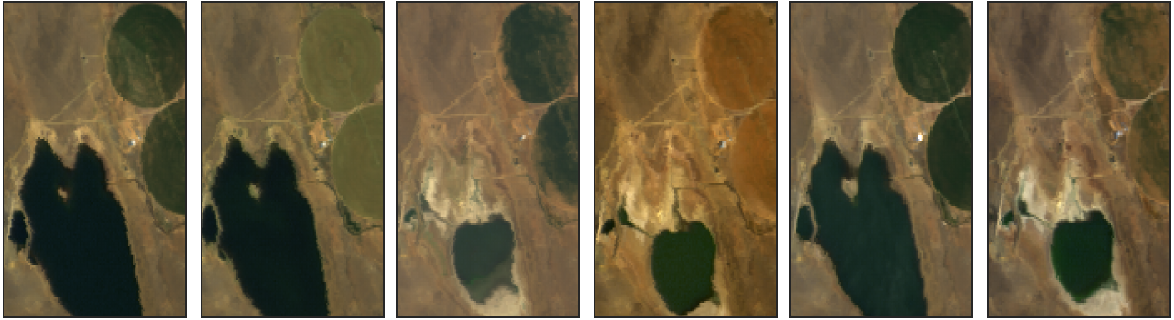}
    \vspace{-0.7cm}
    \caption{True color depiction of the Lake Tahoe HIs and their acquisition dates.}
    \label{fig:illustrative_Tahoe}
\end{figure}

\begin{figure}
    \centering
    {\tiny \hspace{0.05\linewidth} Water \hspace{0.22\linewidth} Soil \hspace{0.2\linewidth} Vegetation} \\
    \vspace{0.2ex}
    \includegraphics[width=0.8\linewidth]{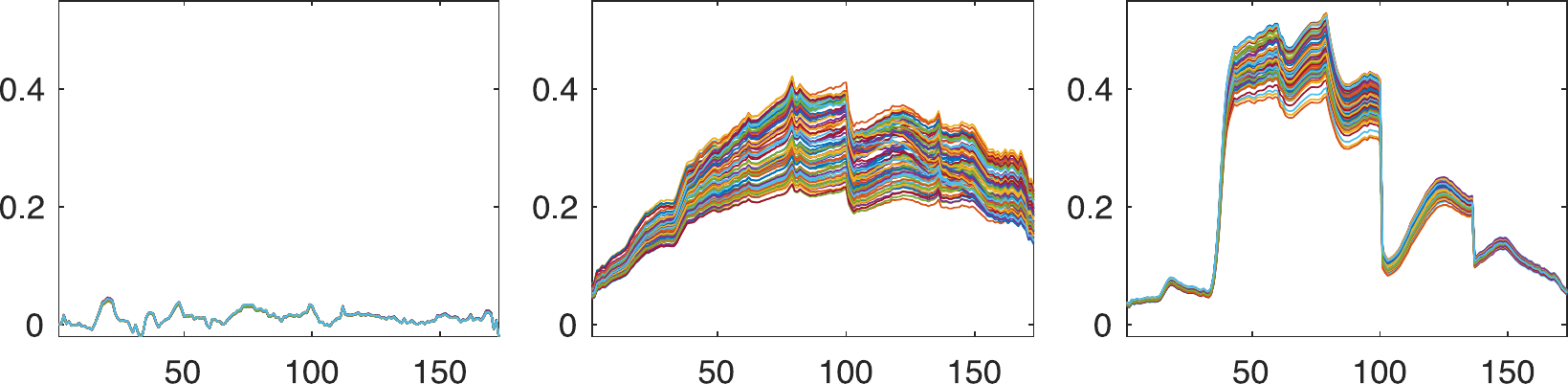}\\ \smallskip
    \includegraphics[width=0.8\linewidth]{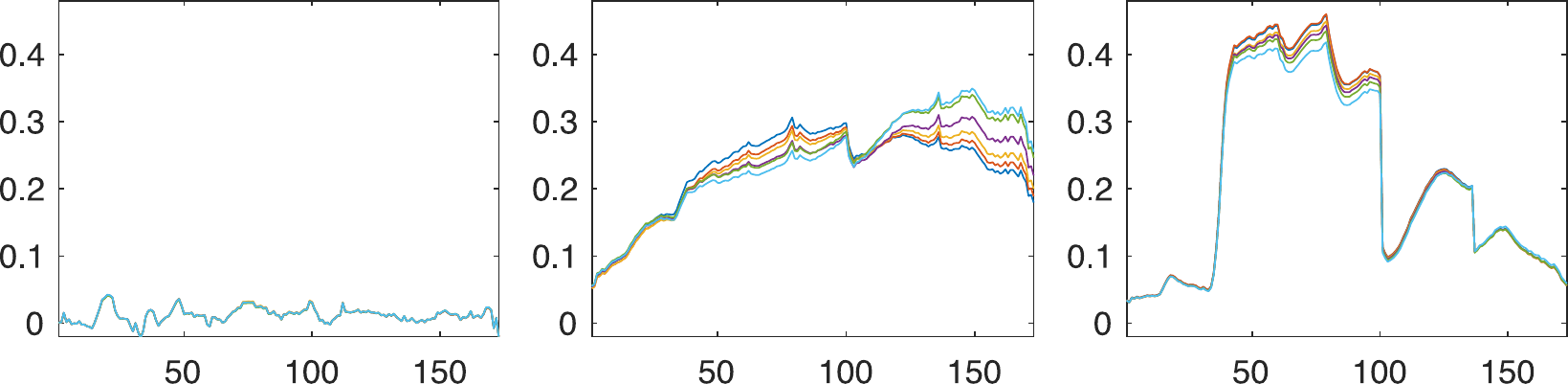}\\
    \vspace{-1.5ex}
    \caption{Estimated EMs for the Lake Tahoe HIs, sampled over space, for time instant $t=3$ (top), and over time, for pixel $n=1$ (bottom).}
    \label{fig:endmembers_Tahoe}
\end{figure}

\begin{figure*}[t]
    \centering
    \hspace{-1.5ex} 
    \includegraphics[width=0.33\linewidth]{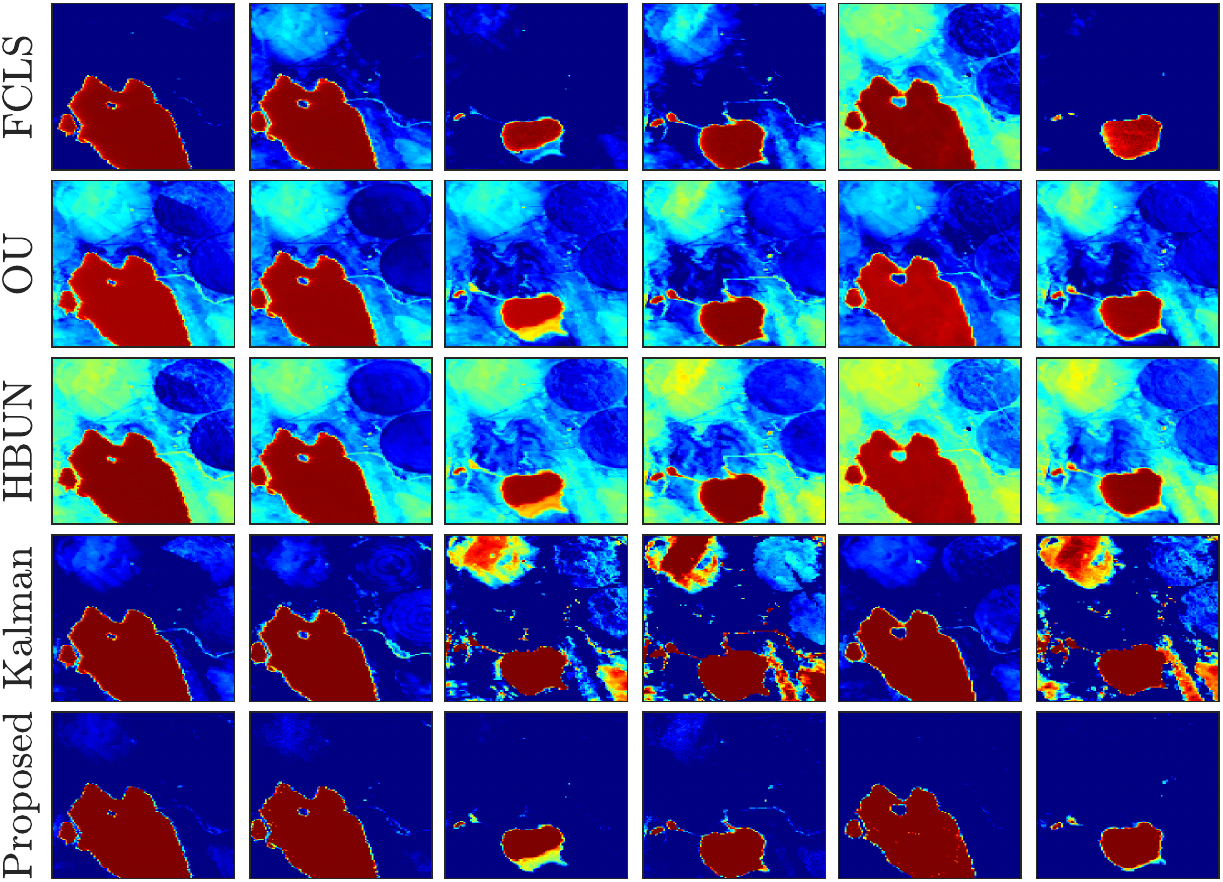}
    \includegraphics[width=0.33\linewidth]{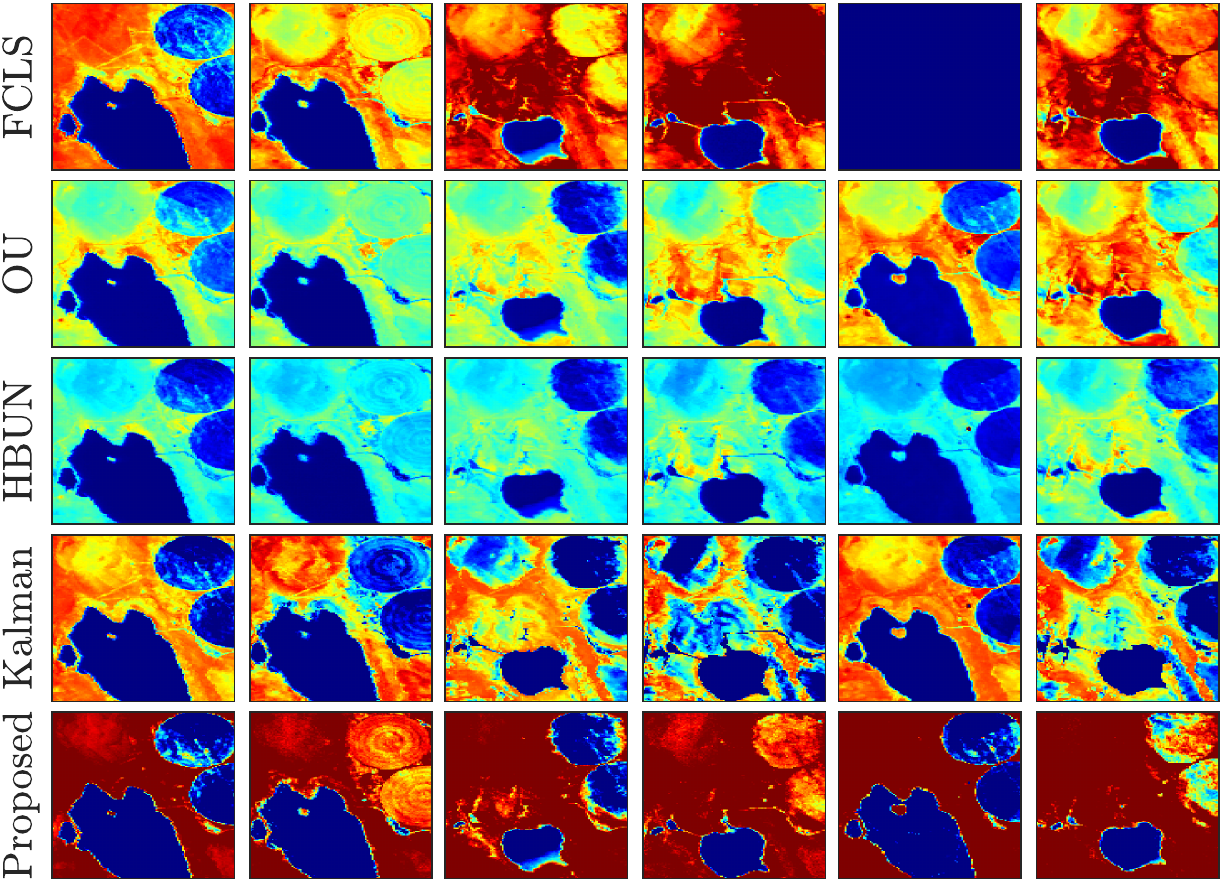}
    \includegraphics[width=0.33\linewidth]{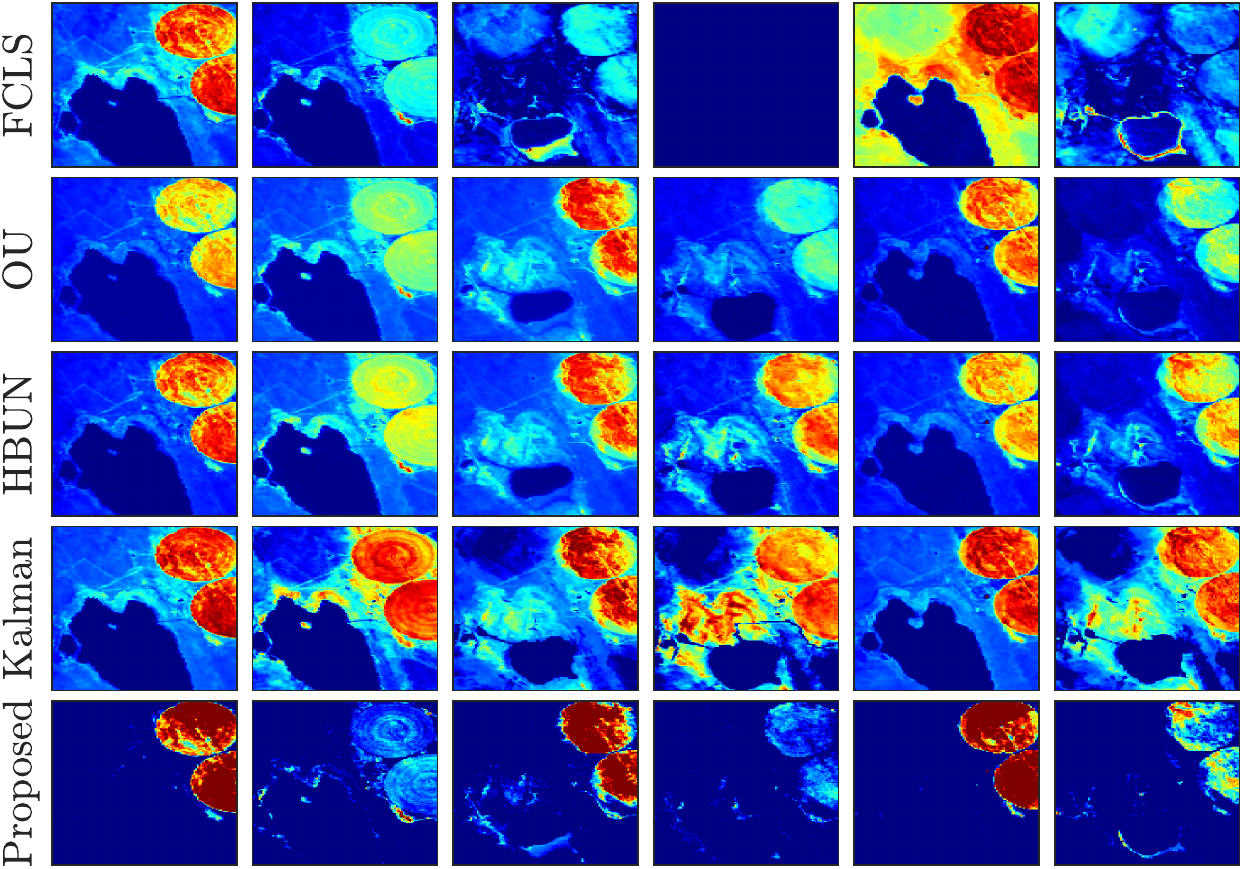} \\
    \vspace{-1.2ex}
    \caption{Estimated abundances for the water (left panel), soil (central panel) and vegetation (right panel) EMs of the Lake Tahoe HIs.}
    \label{fig:tahoe_composite}
\end{figure*}

\subsection{Simulations with real data}
\label{sec:exprtiments_tahoe}

To evaluate the performance of the algorithms on real data, we considered the Lake Tahoe HI sequence, which was originally described in~\cite{thouvenin2016online}.
It consists of a sequence of $T=6$ images acquired over the Lake Tahoe area by the AVIRIS instruments, which are depicted in true color in Figure~\ref{fig:illustrative_Tahoe}. Each HI contained $N=16500$ pixels, and $L=173$ bands were left after the removal of low-SNR and water absorption bands. This scene contains $P=3$ predominant EMs, consisting of soil, water and vegetation, and considerable changes on the lake and on the crop circles can be observed between the images.
The parameters of the ReDSUNN were set as $K=3$, $\sigma_{\psi}=1$, while the parameters of OU, HBUN and of the Kalman filter were selected as described in their original publications.
The recovered abundances are depicted in Figure~\ref{fig:tahoe_composite}, while the recovered EMs (only shown for ReDSUNN due to space limitations) are shown in Figure~\ref{fig:endmembers_Tahoe}. The reconstruction errors and the processing times are presented in Table~\ref{tab:results_realData}.

\subsubsection{Discussion}
From Figure~\ref{fig:tahoe_composite}, it can be seen that the FCLS method did not achieve a good performance in general, particularly for the fifth image in which there was a considerable confusion between the soil and vegetation EMs. The remaining algorithms achieve more stable performance due to taking the temporal information into account.
The OU and the HBUN algorithms (both of which use the PLMM~\cite{Thouvenin_IEEE_TSP_2016_PLMM} model to represent the temporal endmember variability) behaved similarly to each other. Although these methods performed more stably than the FCLS, they still presented considerable water abundances outside of the lake region, which are predominantly composed by soil.
The performance of the Kalman filter method was relatively poor for the third, fourth and sixth images (in which the area of the lake is small), and contained a considerable amount of artifacts. This happens since the Kalman filter method assumes the abundances to be constant over time when estimating the EMs. Consequently, it is not able to handle large abundance changes in the image sequence.
The abundances estimated by ReDSUNN, on the other hand, showed a clear separation between the different materials, adequately capturing the abundance changes occurring in the HIs. Moreover, larger concentrations of mixed pixels were observed in regions that are meaningful, such as at the drying edge of the lake in the third image, and in some parts of the crop circles.
The EMs recovered by ReDSUNN show considerable variability in soil and vegetation spectral, particularly over space, while the water spectra shows little variability. Moreover, the variability of the EMs can be spectrally localized, which can be observed most clearly in the temporal signatures of soil. Moreover, spatial EM variability was more significant than temporal EM variability. Note that the EM variability was also more significant in this example compared to the experiments with synthetic data since a larger value for the hyperparameter $\sigma_{\psi}$ was selected.

The results in Table~\ref{tab:results_realData} show that HBUN and OU obtain the smallest reconstruction errors ($\text{NRMSE}_{\bY}$), while those obtained by Kalman and ReDSUNN, which have less learnable parameters, were larger, with FCLS having the largest $\text{NRMSE}_{\bY}$.
The ratio between the computation times were similar to the synthetic examples, with the Kalman filter being the slowest and the OU the fastest among the MTHU methods that account for temporal information, and the proposed ReDSUNN method achieving intermediate results. This indicates that the methods scale similarly with the image size. Nevertheless, developing more efficient MTHU algorithms is an interesting subject for future work.

\section{Conclusions}
\label{sec:conclusions}

This paper proposed a multitemporal hyperspectral unmixing method based on a variational recurrent neural network. 
A low-dimensional, dynamical state space model was presented to represent the spatial and temporal variations of the endmember spectra by expanding it over a small set of spectrally smooth basis vectors.
The dynamics of the abundances were modelled using a Dirichlet distribution, which was approximated as a Gaussian in the softmax basis in order to improve the efficiency of the inference process.
Based on this generative model, variational inference was considered to perform unmixing by approximating the posterior distribution of the abundances and endmembers.  The Markov and independence properties of the model were also used to improve the efficiency of the solution.
The posterior distribution was parameterized using a combination of a simple, physically interpretable, model and LSTM recurrent neural networks to improve flexibility while maintaining the physical interpretability of the abundances. In the proposed framework, all parameters were computed using stochastic backpropagation.
Experimental results indicate that the proposed algorithm achieves better unmixing performance when compared to state-of-the-art methods, at a similar computational complexity, using both synthetic and real datasets.

\appendices
\section{Computing the terms in~\eqref{eq:pixelntime_factorized_ELBO}}
\label{sec:appendix_KLdivs}

Due to the (conditionally) Gaussian assumptions in the generative model and in the variational posterior, the three terms inside the expectations in~\eqref{eq:pixelntime_factorized_ELBO} can be computed analytically.
The first term in~\eqref{eq:pixelntime_factorized_ELBO} is the log-likelihood of a Gaussian PDF, which can be computed from~\eqref{eq:meas_model_pixelwise_softmax}. The second and third terms are KL divergences between Gaussian PDFs, which can be computed using the general result for two Gaussians of dimension $D$, given by
$\KL\big(\calN(\bmu_1,\bSigma_1) \big\| \calN(\bmu_2,\bSigma_2)\big) = \frac{1}{2} \big(\log\frac{|\bSigma_2|}{|\bSigma_1|} - D
+ \tr\{\bSigma_2^{-1}\bSigma_1\} + (\bmu_2-\bmu_1)^\top \bSigma_2^{-1} (\bmu_2-\bmu_1) \big)$.
The second (resp., third) term are thus computed substituting the mean and covariance from~\eqref{eq:posterior_fact2_init} and~\eqref{eq:final_mdl_init_a},~\eqref{eq:final_mdl_init_psi} (resp.,~\eqref{eq:posterior_fact2} and~\eqref{eq:abundances_time_evol_softmax},~\eqref{eq:dynamical_mdl_psi})
in the expression above by using the fact that $\bc_{n,t}$ and $\bpsi_{n,t}$ are independent in the generative model.
For more details, see, e.g.,~\cite{doersch2016tutorialVAEs}.

\bibliographystyle{IEEEtran}
\bibliography{bibliography_icassp20,references2,references_varRev,references_mtemp,references_VAEu}

\begin{thebibliography}{10}
\providecommand{\url}[1]{#1}
\csname url@samestyle\endcsname
\providecommand{\newblock}{\relax}
\providecommand{\bibinfo}[2]{#2}
\providecommand{\BIBentrySTDinterwordspacing}{\spaceskip=0pt\relax}
\providecommand{\BIBentryALTinterwordstretchfactor}{4}
\providecommand{\BIBentryALTinterwordspacing}{\spaceskip=\fontdimen2\font plus
\BIBentryALTinterwordstretchfactor\fontdimen3\font minus
  \fontdimen4\font\relax}
\providecommand{\BIBforeignlanguage}[2]{{%
\expandafter\ifx\csname l@#1\endcsname\relax
\typeout{** WARNING: IEEEtran.bst: No hyphenation pattern has been}%
\typeout{** loaded for the language `#1'. Using the pattern for}%
\typeout{** the default language instead.}%
\else
\language=\csname l@#1\endcsname
\fi
#2}}
\providecommand{\BIBdecl}{\relax}
\BIBdecl

\bibitem{Bioucas-Dias-2013-ID307}
J.~M. Bioucas-Dias, A.~Plaza, G.~Camps-Valls, P.~Scheunders, N.~Nasrabadi, and
  J.~Chanussot, ``Hyperspectral remote sensing data analysis and future
  challenges,'' \emph{IEEE Geoscience and Remote Sensing Magazine}, vol.~1,
  no.~2, pp. 6--36, 2013.

\bibitem{shaw2003spectralImagRemote}
G.~A. Shaw and H.-h.~K. Burke, ``Spectral imaging for remote sensing,''
  \emph{Lincoln laboratory journal}, vol.~14, no.~1, pp. 3--28, 2003.

\bibitem{Keshava:2002p5667}
N.~Keshava and J.~F. Mustard, ``Spectral unmixing,'' \emph{IEEE Signal
  Processing Magazine}, vol.~19, no.~1, pp. 44--57, 2002.

\bibitem{borsoi2020variabilityReview2}
R.~A. Borsoi, T.~Imbiriba, J.~C.~M. Bermudez, C.~Richard, J.~Chanussot,
  L.~Drumetz, J.-Y. Tourneret, A.~Zare, and C.~Jutten, ``Spectral variability
  in hyperspectral data unmixing: A comprehensive review,'' \emph{IEEE
  Geoscience and Remote Sensing Magazine}, vol.~9, pp. 223--270, 2021.

\bibitem{Zare-2014-ID324-variabilityReview}
A.~Zare and K.~C. Ho, ``Endmember variability in hyperspectral analysis:
  Addressing spectral variability during spectral unmixing,'' \emph{IEEE Signal
  Processing Magazine}, vol.~31, pp. 95--104, January 2014.

\bibitem{henrot2016dynamical}
S.~Henrot, J.~Chanussot, and C.~Jutten, ``Dynamical spectral unmixing of
  multitemporal hyperspectral images,'' \emph{IEEE Trans. Image Process.},
  vol.~25, no.~7, pp. 3219--3232, 2016.

\bibitem{thouvenin2016online}
P.-A. Thouvenin, N.~Dobigeon, and J.-Y. Tourneret, ``Online unmixing of
  multitemporal hyperspectral images accounting for spectral variability,''
  \emph{IEEE Trans. Image Process.}, vol.~25, no.~9, pp. 3979--3990, 2016.

\bibitem{borsoi2020multitemporalUKalmanEM}
R.~A. Borsoi, T.~Imbiriba, P.~Closas, J.~C.~M. Bermudez, and C.~Richard,
  ``Kalman filtering and expectation maximization for multitemporal spectral
  unmixing,'' \emph{IEEE Geoscience and Remote Sensing Letters}, 2020.

\bibitem{liu2021bayesianSU_multitemporal}
H.~Liu, Y.~Lu, Z.~Wu, Q.~Du, J.~Chanussot, and Z.~Wei, ``Bayesian unmixing of
  hyperspectral image sequence with composite priors for abundance and
  endmember variability,'' \emph{IEEE Transactions on Geoscience and Remote
  Sensing}, vol.~60, pp. 1--15, 2021.

\bibitem{somers2013invasiveHawaiiMultiTemporalBandWeighting}
B.~Somers and G.~P. Asner, ``Invasive species mapping in hawaiian rainforests
  using multi-temporal hyperion spaceborne imaging spectroscopy,'' \emph{IEEE
  Journal of Selected Topics in Applied Earth Observations and Remote Sensing},
  vol.~6, no.~2, pp. 351--359, 2013.

\bibitem{somers2013uncorrelatedBandSelectionInstabilityIndex}
------, ``Multi-temporal hyperspectral mixture analysis and feature selection
  for invasive species mapping in rainforests,'' \emph{Remote Sensing of
  Environment}, vol. 136, pp. 14--27, 2013.

\bibitem{lippitt2018multidateMESMAshrublands}
C.~L. Lippitt, D.~A. Stow, D.~A. Roberts, and L.~L. Coulter, ``Multidate
  {MESMA} for monitoring vegetation growth forms in southern california
  shrublands,'' \emph{International journal of remote sensing}, vol.~39, no.~3,
  pp. 655--683, 2018.

\bibitem{goenaga2013unmixingTimeSeriesPuertoRico}
M.~A. Goenaga, M.~C. Torres-Madronero, M.~Velez-Reyes, S.~J. Van~Bloem, and
  J.~D. Chinea, ``Unmixing analysis of a time series of hyperion images over
  the gu{\'a}nica dry forest in puerto rico,'' \emph{IEEE Journal of Selected
  Topics in Applied Earth Observations and Remote Sensing}, vol.~6, no.~2, pp.
  329--338, 2013.

\bibitem{liu2019reviewCD}
S.~Liu, D.~Marinelli, L.~Bruzzone, and F.~Bovolo, ``A review of change
  detection in multitemporal hyperspectral images: Current techniques,
  applications, and challenges,'' \emph{IEEE Geoscience and Remote Sensing
  Magazine}, vol.~7, no.~2, pp. 140--158, 2019.

\bibitem{guo2021changeDetUnmixing}
Q.~Guo, J.~Zhang, C.~Zhong, and Y.~Zhang, ``Change detection for hyperspectral
  images via convolutional sparse analysis and temporal spectral unmixing,''
  \emph{IEEE Journal of Selected Topics in Applied Earth Observations and
  Remote Sensing}, vol.~14, pp. 4417--4426, 2021.

\bibitem{borsoi2021MT_MESMA}
R.~A. Borsoi, T.~Imbiriba, J.~C.~M. Bermudez, and C.~Richard, ``Fast unmixing
  and change detection in multitemporal hyperspectral data,'' \emph{IEEE
  Transactions on Computational Imaging}, vol.~7, pp. 975--988, 2021.

\bibitem{bhatt2020deepLearningHUreview}
J.~S. Bhatt and M.~V. Joshi, ``Deep learning in hyperspectral unmixing: A
  review,'' in \emph{Proc. IEEE International Geoscience and Remote Sensing
  Symposium (IGARSS)}.\hskip 1em plus 0.5em minus 0.4em\relax IEEE, 2020, pp.
  2189--2192.

\bibitem{palsson2022unmixingAECcomparison}
B.~Palsson, J.~R. Sveinsson, and M.~O. Ulfarsson, ``Blind hyperspectral
  unmixing using autoencoders: A critical comparison,'' \emph{IEEE Journal of
  Selected Topics in Applied Earth Observations and Remote Sensing}, vol.~15,
  pp. 1340--1372, 2022.

\bibitem{zhou2021ADMM_SU_networks}
C.~Zhou and M.~R. Rodrigues, ``{ADMM-Based} hyperspectral unmixing networks for
  abundance and endmember estimation,'' \emph{IEEE Transactions on Geoscience
  and Remote Sensing}, 2021.

\bibitem{wang2019AECnlin}
M.~Wang, M.~Zhao, J.~Chen, and S.~Rahardja, ``Nonlinear unmixing of
  hyperspectral data via deep autoencoder networks,'' \emph{IEEE Geoscience and
  Remote Sensing Letters}, vol.~16, no.~9, pp. 1467--1471, 2019.

\bibitem{borsoi2019deepGun}
R.~A. Borsoi, T.~Imbiriba, and J.~C.~M. Bermudez, ``Deep generative endmember
  modeling: {An} application to unsupervised spectral unmixing,'' \emph{IEEE
  Transactions on Computational Imaging}, vol.~6, pp. 374--384, 2019.

\bibitem{imbiriba2022hybrid}
T.~Imbiriba, A.~Demirkaya, J.~Dun{\'\i}k, O.~Straka, D.~Erdo{\u{g}}mu{\c{s}},
  and P.~Closas, ``{Hybrid Neural Network Augmented Physics-based Models for
  Nonlinear Filtering},'' in \emph{Proc. FUSION conference}, Linkoping, Sweden,
  2022.

\bibitem{barber2012ML_book}
D.~Barber, \emph{{Bayesian Reasoning and Machine Learning}}.\hskip 1em plus
  0.5em minus 0.4em\relax {Cambridge University Press}, 2012.

\bibitem{kent1988spatialClassificationFuzzy}
J.~T. Kent and K.~V. Mardia, ``Spatial classification using fuzzy membership
  models,'' \emph{IEEE Transactions on Pattern Analysis and Machine
  Intelligence}, vol.~10, no.~5, pp. 659--671, 1988.

\bibitem{eches2011bayesianSpatialMarkovUnmixing}
O.~Eches, N.~Dobigeon, and J.-Y. Tourneret, ``Enhancing hyperspectral image
  unmixing with spatial correlations,'' \emph{IEEE Transactions on Geoscience
  and Remote Sensing}, vol.~49, no.~11, pp. 4239--4247, 2011.

\bibitem{mackay1998choiceBasisDirichlet}
D.~J. MacKay, ``Choice of basis for {Laplace} approximation,'' \emph{Machine
  learning}, vol.~33, no.~1, pp. 77--86, 1998.

\bibitem{kingma14VAEs}
D.~P. Kingma and M.~Welling, ``Auto-encoding variational bayes,'' in
  \emph{Proc. 2nd International Conference on Learning Representations (ICLR)},
  Y.~Bengio and Y.~LeCun, Eds., Banff, AB, Canada, 2014.

\bibitem{lipton2015criticalReviewRNNs}
Z.~C. Lipton, J.~Berkowitz, and C.~Elkan, ``A critical review of recurrent
  neural networks for sequence learning,'' \emph{arXiv preprint
  arXiv:1506.00019}, 2015.

\bibitem{sigurdsson2017sparseDistU}
J.~Sigurdsson, M.~O. Ulfarsson, J.~R. Sveinsson, and J.~Bioucas-Dias, ``Sparse
  distributed multitemporal hyperspectral unmixing,'' \emph{IEEE Trans. Geosci.
  Remote Sens.}, vol.~55, no.~11, pp. 6069--6084, 2017.

\bibitem{thouvenin2018hierarchicalBU}
P.-A. Thouvenin, N.~Dobigeon, and J.-Y. Tourneret, ``A hierarchical bayesian
  model accounting for endmember variability and abrupt spectral changes to
  unmix multitemporal hyperspectral images,'' \emph{IEEE Trans. Comput.
  Imaging}, vol.~4, no.~1, pp. 32--45, 2018.

\bibitem{zhuo2022spectralTemporalBayesianSUSentinel2}
R.~Zhuo, Y.~Fang, L.~Xu, Y.~Chen, Y.~Wang, and J.~Peng, ``A novel
  spectral-temporal {Bayesian} unmixing algorithm with spatial prior for
  {Sentinel-2} time series,'' \emph{Remote Sensing Letters}, vol.~13, no.~5,
  pp. 522--532, 2022.

\bibitem{roberts1998originalMESMA}
D.~A. Roberts, M.~Gardner, R.~Church, S.~Ustin, G.~Scheer, and R.~O. Green,
  ``Mapping chaparral in the santa monica mountains using multiple endmember
  spectral mixture models,'' \emph{Remote Sensing of Environment}, vol.~65,
  no.~3, pp. 267--279, 1998.

\bibitem{dudley2015multitemporalLibraryPhenologicalGradiantsMESMA}
K.~L. Dudley, P.~E. Dennison, K.~L. Roth, D.~A. Roberts, and A.~R. Coates, ``A
  multi-temporal spectral library approach for mapping vegetation species
  across spatial and temporal phenological gradients,'' \emph{Remote Sensing of
  Environment}, vol. 167, pp. 121--134, 2015.

\bibitem{wang2021spatioTemporalSUtimeseries}
Q.~Wang, X.~Ding, X.~Tong, and P.~M. Atkinson, ``Spatio-temporal spectral
  unmixing of time-series images,'' \emph{Remote Sensing of Environment}, vol.
  259, p. 112407, 2021.

\bibitem{wang2021realTimeSU_MODIS}
------, ``Real-time spatiotemporal spectral unmixing of {MODIS} images,''
  \emph{IEEE Transactions on Geoscience and Remote Sensing}, 2021.

\bibitem{somers2011variabilityReview}
B.~Somers, G.~P. Asner, L.~Tits, and P.~Coppin, ``Endmember variability in
  spectral mixture analysis: A review,'' \emph{Remote Sensing of Environment},
  vol. 115, no.~7, pp. 1603--1616, 2011.

\bibitem{iordache2011sunsal}
M.-D. Iordache, J.~M. Bioucas-Dias, and A.~Plaza, ``Sparse unmixing of
  hyperspectral data,'' \emph{IEEE Transactions on Geoscience and Remote
  Sensing}, vol.~49, no.~6, pp. 2014--2039, 2011.

\bibitem{borsoi2018superpixels1_sparseU}
R.~A. {Borsoi}, T.~{Imbiriba}, J.~C.~M. {Bermudez}, and C.~{Richard}, ``A fast
  multiscale spatial regularization for sparse hyperspectral unmixing,''
  \emph{IEEE Geoscience and Remote Sensing Letters}, vol.~16, no.~4, pp.
  598--602, April 2019.

\bibitem{drumetz2019SU_bundlesGroupSparsityMixedNorms}
L.~Drumetz, T.~R. Meyer, J.~Chanussot, A.~L. Bertozzi, and C.~Jutten,
  ``Hyperspectral image unmixing with endmember bundles and group sparsity
  inducing mixed norms,'' \emph{IEEE Transactions on Image Processing},
  vol.~28, no.~7, pp. 3435--3450, 2019.

\bibitem{uezato2018SU_variabilityAdaptiveBundlesDoubleSparse}
T.~Uezato, M.~Fauvel, and N.~Dobigeon, ``Hyperspectral unmixing with spectral
  variability using adaptive bundles and double sparsity,'' \emph{IEEE
  Transactions on Geoscience and Remote Sensing}, vol.~57, no.~6, pp.
  3980--3992, 2019.

\bibitem{halimi2015unsupervisedBayesianUnmixing}
A.~Halimi, N.~Dobigeon, and J.-Y. Tourneret, ``Unsupervised unmixing of
  hyperspectral images accounting for endmember variability,'' \emph{IEEE
  Transactions on Image Processing}, vol.~24, no.~12, pp. 4904--4917, 2015.

\bibitem{zhou2018variabilityGaussianMixtureModel}
Y.~Zhou, A.~Rangarajan, and P.~D. Gader, ``A gaussian mixture model
  representation of endmember variability in hyperspectral unmixing,''
  \emph{IEEE Transactions on Image Processing}, vol.~27, no.~5, pp. 2242--2256,
  May 2018.

\bibitem{du2014spatialBetaCompositional}
X.~Du, A.~Zare, P.~Gader, and D.~Dranishnikov, ``Spatial and spectral unmixing
  using the beta compositional model,'' \emph{IEEE Journal of Selected Topics
  in Applied Earth Observations and Remote Sensing}, vol.~7, no.~6, pp.
  1994--2003, 2014.

\bibitem{Thouvenin_IEEE_TSP_2016_PLMM}
P.-A. Thouvenin, N.~Dobigeon, and J.-Y. Tourneret, ``Hyperspectral unmixing
  with spectral variability using a perturbed linear mixing model,'' \emph{IEEE
  Trans. Signal Processing}, vol.~64, no.~2, pp. 525--538, Feb. 2016.

\bibitem{drumetz2016blindUnmixingELMM}
L.~Drumetz, M.-A. Veganzones, S.~Henrot, R.~Phlypo, J.~Chanussot, and
  C.~Jutten, ``Blind hyperspectral unmixing using an extended linear mixing
  model to address spectral variability,'' \emph{IEEE Trans. Image Process.},
  vol.~25, no.~8, pp. 3890--3905, 2016.

\bibitem{imbiriba2018GLMM}
T.~Imbiriba, R.~A. Borsoi, and J.~C.~M. Bermudez, ``Generalized linear mixing
  model accounting for endmember variability,'' in \emph{Proc. IEEE ICASSP},
  Calgary, Canada, 2018, pp. 1862--1866.

\bibitem{hong2019augmentedLMMvariability}
D.~Hong, N.~Yokoya, J.~Chanussot, and X.~X. Zhu, ``An augmented linear mixing
  model to address spectral variability for hyperspectral unmixing,''
  \emph{IEEE Transactions on Image Processing}, vol.~28, no.~4, pp. 1923--1938,
  2019.

\bibitem{Borsoi_multiscaleVar_2018_TIP}
R.~A. {Borsoi}, T.~{Imbiriba}, and J.~C. {Moreira Bermudez}, ``A data dependent
  multiscale model for hyperspectral unmixing with spectral variability,''
  \emph{IEEE Transactions on Image Processing}, vol.~29, pp. 3638--3651, 2020.

\bibitem{imbiriba2018ULTRA_V}
T.~Imbiriba, R.~A. Borsoi, and J.~C.~M. Bermudez, ``Low-rank tensor modeling
  for hyperspectral unmixing accounting for spectral variability,'' \emph{IEEE
  Transactions on Geoscience and Remote Sensing}, vol.~58, no.~3, pp.
  1833--1842, 2020.

\bibitem{uezato2020illuminationSU_surfaceModel}
T.~Uezato, N.~Yokoya, and W.~He, ``Illumination invariant hyperspectral image
  unmixing based on a digital surface model,'' \emph{IEEE Transactions on Image
  Processing}, vol.~29, pp. 3652--3664, 2020.

\bibitem{guilfoyle2001comparativeUnmixingNeuralNetworksRBF}
K.~J. Guilfoyle, M.~L. Althouse, and C.-I. Chang, ``A quantitative and
  comparative analysis of linear and nonlinear spectral mixture models using
  radial basis function neural networks,'' \emph{IEEE Transactions on
  Geoscience and Remote Sensing}, vol.~39, no.~10, pp. 2314--2318, 2001.

\bibitem{plaza2010selectingTrainingSamplesNNunmixing}
J.~Plaza and A.~Plaza, ``Spectral mixture analysis of hyperspectral scenes
  using intelligently selected training samples,'' \emph{IEEE Geoscience and
  Remote Sensing Letters}, vol.~7, no.~2, pp. 371--375, 2010.

\bibitem{guo2015autoencodersUnmixing}
R.~Guo, W.~Wang, and H.~Qi, ``Hyperspectral image unmixing using autoencoder
  cascade,'' in \emph{Proc. 7th Workshop on Hyperspectral Image and Signal
  Processing: Evolution in Remote Sensing (WHISPERS)}, Tokyo, Japan, June 2015,
  pp. 1--4.

\bibitem{palsson2018autoencoderUnmixing_IEEEaccess}
B.~Palsson, J.~Sigurdsson, J.~R. Sveinsson, and M.~O. Ulfarsson,
  ``Hyperspectral unmixing using a neural network autoencoder,'' \emph{IEEE
  Access}, vol.~6, pp. 25\,646--25\,656, 2018.

\bibitem{su2019deepAutoencoderUnmixing}
Y.~Su, J.~Li, A.~Plaza, A.~Marinoni, P.~Gamba, and S.~Chakravortty, ``{DAEN}:
  Deep autoencoder networks for hyperspectral unmixing,'' \emph{IEEE
  Transactions on Geoscience and Remote Sensing}, vol.~57, no.~7, pp.
  4309--4321, 2019.

\bibitem{sahoo2022HU_geological_latentEncoding}
M.~M. Sahoo, A.~Porwal, A.~Karnieli \emph{et~al.}, ``Deep-learning-based latent
  space encoding for spectral unmixing of geological materials,'' \emph{ISPRS
  Journal of Photogrammetry and Remote Sensing}, vol. 183, pp. 307--320, 2022.

\bibitem{palsson2020convolutionalAEC_SU}
B.~Palsson, M.~O. Ulfarsson, and J.~R. Sveinsson, ``Convolutional autoencoder
  for spectral-spatial hyperspectral unmixing,'' \emph{IEEE Transactions on
  Geoscience and Remote Sensing}, pp. 1--15, 2020.

\bibitem{hua2021gatedAEC_SU}
Z.~Hua, X.~Li, J.~Jiang, and L.~Zhao, ``Gated autoencoder network for
  spectral--spatial hyperspectral unmixing,'' \emph{Remote Sensing}, vol.~13,
  no.~16, p. 3147, 2021.

\bibitem{qu2018udas_autoencoderUnmixing}
Y.~Qu and H.~Qi, ``{uDAS}: An untied denoising autoencoder with sparsity for
  spectral unmixing,'' \emph{IEEE Transactions on Geoscience and Remote
  Sensing}, vol.~57, no.~3, pp. 1698--1712, March 2019.

\bibitem{zhao2021AECnonlinearSUattitive3D}
M.~Zhao, M.~Wang, J.~Chen, and S.~Rahardja, ``Hyperspectral unmixing for
  additive nonlinear models with a {3-D-CNN} autoencoder network,'' \emph{IEEE
  Transactions on Geoscience and Remote Sensing}, 2021.

\bibitem{shahid2021unsupervisedSUautoencoder}
K.~T. Shahid and I.~D. Schizas, ``Unsupervised hyperspectral unmixing via
  nonlinear autoencoders,'' \emph{IEEE Transactions on Geoscience and Remote
  Sensing}, 2021.

\bibitem{li2021modelBasedAECsSU}
H.~Li, R.~A. Borsoi, T.~Imbiriba, P.~Closas, J.~C. Bermudez, and
  D.~Erdo{\u{g}}mu{\c{s}}, ``Model-based deep autoencoder networks for
  nonlinear hyperspectral unmixing,'' \emph{IEEE Geoscience and Remote Sensing
  Letters}, vol.~19, pp. 1--5, 2021.

\bibitem{borsoi2019EMlibManInterpVAE}
R.~A. Borsoi, T.~Imbiriba, J.~C.~M. Bermudez, and C.~Richard, ``Deep generative
  models for library augmentation in multiple endmember spectral mixture
  analysis,'' \emph{IEEE Geoscience and Remote Sensing Letters}, 2020.

\bibitem{shi2021generativeModelEMvariability}
S.~Shi, M.~Zhao, L.~Zhang, Y.~Altmann, and J.~Chen, ``Probabilistic generative
  model for hyperspectral unmixing accounting for endmember variability,''
  \emph{IEEE Transactions on Geoscience and Remote Sensing}, 2021.

\bibitem{uezato2016unmixingGaussianProcessVariability}
T.~Uezato, R.~J. Murphy, A.~Melkumyan, and A.~Chlingaryan, ``A novel spectral
  unmixing method incorporating spectral variability within endmember
  classes,'' \emph{IEEE Transactions on Geoscience and Remote Sensing},
  vol.~54, no.~5, pp. 2812--2831, 2016.

\bibitem{koirala2020geodesicSupervisedSUvariability}
B.~Koirala, Z.~Zahiri, A.~Lamberti, and P.~Scheunders, ``Robust supervised
  method for nonlinear spectral unmixing accounting for endmember
  variability,'' \emph{IEEE Transactions on Geoscience and Remote Sensing},
  2020, doi:~\url{10.1109/TGRS.2020.3031012}.

\bibitem{jin2021two_stream_AEC_SU}
Q.~Jin, Y.~Ma, X.~Mei, and J.~Ma, ``Tanet: An unsupervised two-stream
  autoencoder network for hyperspectral unmixing,'' \emph{IEEE Transactions on
  Geoscience and Remote Sensing}, 2021.

\bibitem{hong2021egu_net}
D.~Hong, L.~Gao, J.~Yao, N.~Yokoya, J.~Chanussot, U.~Heiden, and B.~Zhang,
  ``Endmember-guided unmixing network {(EGU-Net)}: A general deep learning
  framework for self-supervised hyperspectral unmixing,'' \emph{IEEE
  Transactions on Neural Networks and Learning Systems}, 2021.

\bibitem{li2021selfSuperv_deepNMF_SU}
H.-C. Li, X.-R. Feng, D.-H. Zhai, Q.~Du, and A.~Plaza, ``Self-supervised robust
  deep matrix factorization for hyperspectral unmixing,'' \emph{IEEE
  Transactions on Geoscience and Remote Sensing}, 2021.

\bibitem{drumetz2020learningEndmemberDynamics}
L.~Drumetz, M.~Dalla~Mura, G.~Tochon, and R.~Fablet, ``Learning endmember
  dynamics in multitemporal hyperspectral data using a state-space model
  formulation,'' in \emph{IEEE International Conference on Acoustics, Speech
  and Signal Processing (ICASSP)}.\hskip 1em plus 0.5em minus 0.4em\relax IEEE,
  2020, pp. 2483--2487.

\bibitem{min2021metricLearningNet_SU}
A.~Min, Z.~Guo, H.~Li, and J.~Peng, ``{JMnet}: Joint metric neural network for
  hyperspectral unmixing,'' \emph{IEEE Transactions on Geoscience and Remote
  Sensing}, vol.~60, pp. 1--12, 2021.

\bibitem{jin2021adversarialAEC_SU}
Q.~Jin, Y.~Ma, F.~Fan, J.~Huang, X.~Mei, and J.~Ma, ``Adversarial autoencoder
  network for hyperspectral unmixing,'' \emph{IEEE Transactions on Neural
  Networks and Learning Systems}, 2021.

\bibitem{gao2021cycu_neT_SU}
L.~Gao, Z.~Han, D.~Hong, B.~Zhang, and J.~Chanussot, ``{CyCU-Net}:
  Cycle-consistency unmixing network by learning cascaded autoencoders,''
  \emph{IEEE Transactions on Geoscience and Remote Sensing}, vol.~60, pp.
  1--14, 2021.

\bibitem{rasti2021HU_deepImagePrior}
B.~Rasti, B.~Koirala, P.~Scheunders, and P.~Ghamisi, ``{UnDIP}: Hyperspectral
  unmixing using deep image prior,'' \emph{IEEE Transactions on Geoscience and
  Remote Sensing}, vol.~60, pp. 1--15, 2021.

\bibitem{sarkka2013bayesian}
S.~S{\"a}rkk{\"a}, \emph{Bayesian filtering and smoothing}.\hskip 1em plus
  0.5em minus 0.4em\relax Cambridge University Press, 2013, vol.~3.

\bibitem{murphy2012machineLearningBook}
K.~P. Murphy, \emph{Machine learning: a probabilistic perspective}.\hskip 1em
  plus 0.5em minus 0.4em\relax MIT press, 2012.

\bibitem{Borsoi_2018_Fusion}
R.~A. {Borsoi}, T.~{Imbiriba}, and J.~C.~M. {Bermudez}, ``Super-resolution for
  hyperspectral and multispectral image fusion accounting for seasonal spectral
  variability,'' \emph{IEEE Trans. Image Process.}, vol.~29, no.~1, pp.
  116--127, 2020.

\bibitem{halimi2016unmixingVariabilityNonlinearityMismodeling}
A.~Halimi, P.~Honeine, and J.~M. Bioucas-Dias, ``Hyperspectral unmixing in
  presence of endmember variability, nonlinearity, or mismodeling effects,''
  \emph{IEEE Transactions on Image Processing}, vol.~25, no.~10, pp.
  4565--4579, 2016.

\bibitem{Halimi_IEEE_Trans_CI_2017}
A.~Halimi, J.~Bioucas-Dias, N.~Dobigeon, G.~S. Buller, and S.~McLaughlin,
  ``Fast hyperspectral unmixing in presence of nonlinearity or mismodelling
  effects,'' \emph{IEEE Trans. Computational Imaging}, vol.~3, no.~2, pp.
  146--159, April 2017.

\bibitem{Hapke1981}
B.~Hapke, ``{Bidirectional reflectance spectroscopy, 1, Theory},''
  \emph{Journal of Geophysical Research}, vol.~86, no.~B4, pp. 3039--3054,
  1981.

\bibitem{jacquemoud2001leafOpticalPropertiesReview}
S.~Jacquemoud and S.~L. Ustin, ``Leaf optical properties: A state of the art,''
  in \emph{8th International Symposium of Physical Measurements \& Signatures
  in Remote Sensing}.\hskip 1em plus 0.5em minus 0.4em\relax CNES, Aussois
  France, 2001, pp. 223--332.

\bibitem{Halimi_IEEE_TIP_2015}
A.~Halimi, N.~Dobigeon, and J.-Y. Tourneret, ``Unsupervised unmixing of
  hyperspectral images accounting for endmember variability,'' \emph{IEEE
  Trans. Image Processing}, vol.~24, no.~12, pp. 4904--4917, Dec. 2015.

\bibitem{eches2012adaptiveMRFunmixing}
O.~Eches, J.~A. Benediktsson, N.~Dobigeon, and J.-Y. Tourneret, ``Adaptive
  {Markov} random fields for joint unmixing and segmentation of hyperspectral
  images,'' \emph{IEEE Transactions on Image Processing}, vol.~22, no.~1, pp.
  5--16, 2012.

\bibitem{hennig2012kernelTopicModels}
P.~Hennig, D.~Stern, R.~Herbrich, and T.~Graepel, ``Kernel topic models,'' in
  \emph{Artificial intelligence and statistics}.\hskip 1em plus 0.5em minus
  0.4em\relax PMLR, 2012, pp. 511--519.

\bibitem{srivastava2017autoencodingTopicModels}
A.~Srivastava and C.~Sutton, ``Autoencoding variational inference for topic
  models,'' \emph{arXiv preprint arXiv:1703.01488}, 2017.

\bibitem{kingma2014VAEs}
D.~P. Kingma and M.~Welling, ``Auto-encoding variational bayes,'' in
  \emph{Proc. 2nd International Conference on Learning Representations (ICLR)},
  Y.~Bengio and Y.~LeCun, Eds., Banff, AB, Canada, April 14-16, 2014.

\bibitem{kingma2014adam}
D.~P. Kingma and J.~Ba, ``Adam: A method for stochastic optimization,'' in
  \emph{Proc. International Conf. on Learning Representations (ICLR)}, 2015.

\bibitem{krishnan2017structuredInferenceNets}
R.~G. Krishnan, U.~Shalit, and D.~Sontag, ``Structured inference networks for
  nonlinear state space models,'' in \emph{Proceedings of the Thirty-First AAAI
  Conference on Artificial Intelligence}, 2017, pp. 2101--2109.

\bibitem{hochreiter1997lstm}
S.~Hochreiter and J.~Schmidhuber, ``Long short-term memory,'' \emph{Neural
  computation}, vol.~9, no.~8, pp. 1735--1780, 1997.

\bibitem{glorot2010understandingInitializeNNs}
X.~Glorot and Y.~Bengio, ``Understanding the difficulty of training deep
  feedforward neural networks,'' in \emph{Proceedings of the thirteenth
  international conference on artificial intelligence and statistics}.\hskip
  1em plus 0.5em minus 0.4em\relax JMLR Workshop and Conference Proceedings,
  2010, pp. 249--256.

\bibitem{Nascimento2005}
J.~M.~P. Nascimento and J.~M. Bioucas-Dias, ``{Vertex Component Analysis}: A
  fast algorithm to unmix hyperspectral data,'' \emph{IEEE Trans. Geosci.
  Remote Sens.}, vol.~43, no.~4, pp. 898--910, April 2005.

\bibitem{doersch2016tutorialVAEs}
C.~Doersch, ``Tutorial on variational autoencoders,'' \emph{arXiv preprint
  arXiv:1606.05908}, 2016.

\end{thebibliography}

\end{document}